\documentclass[preprint,superscriptaddress,longbibliography]{revtex4-1}
\usepackage{graphicx}
\usepackage{epstopdf, epsfig}
\usepackage{hyperref}
\usepackage[english]{babel}
\usepackage{amssymb}
\usepackage{subfigure}
\usepackage{xparse}

\newcounter{subfigs}[figure]

\makeatletter
\NewDocumentCommand{\subfigimg}{s O{} m D<>{10pt} O{2\baselineskip} m}{%
    \IfBooleanTF{#1}%
        {\@subfigimg{#2}{#3}{#6}{0}{#4}{#5}}%
        {\@subfigimg{#2}{#3}{#6}{1}{#4}{#5}}%
}
\newcommand*{\@subfigimg}[6]{%
  \bgroup%
  \advance\c@figure by #4%
  \refstepcounter{subfigs}%
  \ifx\relax#3\relax\else%
  \label{#3}%
  \fi%
  \setbox1=\hbox{\includegraphics[#1]{#2}}
  \leavevmode\rlap{\usebox1}
  \rlap{\hspace*{-7.5pt}\raisebox{\dimexpr\ht1-#6\relax}{(\alph{subfigs})}}
  \phantom{\usebox1}
  \egroup%
}

\makeatother
\usepackage{xr}
\usepackage{amsmath}
\usepackage{xcolor}
\usepackage{epstopdf}
\usepackage{color}
\usepackage{ulem}
\usepackage{multirow}

\begin{document}

\title{Isochrons, Phase Response and Synchronization Dynamics \\
of Tunable Photonic Oscillators}

\author{Georgia Himona}
\affiliation{School of Applied Mathematical and Physical Sciences, National Technical University of Athens, Athens 15780, Greece}

\author{Vassilios Kovanis}
\affiliation{Bradley Department of Electrical and Computer Engineering, Virginia Tech Research Center in Arlington Virginia, Arlington Virginia 22203, USA}

\author{Yannis Kominis}
\email[corresponding author (email): ]{gkomin@central.ntua.gr}
\affiliation{School of Applied Mathematical and Physical Sciences, National Technical University of Athens, Athens 15780, Greece}

\begin{abstract}
The global structure of the Isochrons and the corresponding Phase Response Curves are, for the first time, investigated and numerically computed for the fundamental photonic oscillator consisted of an Optically Injected Laser. Their crucial role in the synchronization dynamics under a periodic modulation of the injection beam is shown, along with their capability for providing conditions for stable phase-locking and periodic outputs with discrete equidistant spectra in the form of frequency combs.  
\end{abstract}

\maketitle

Optically Injected Lasers (OILs) are widely used tunable photonic oscillators of fundamental technological importance, that have been studied, both theoretically and experimentally, for more than four decades. They are well known for a remarkably rich set of complex dynamical features such as different types of instabilities, cascades of bifurcations, multi-stability, and chaotic transitions \cite{Erneux_book,Phys_Rep, Ohtsubo_13}. Among them, the existence of self-sustained oscillations corresponding to stable limit cycles (LC) has the most fundamental role in terms of their practical applications as tunable photonic oscillators. Their dynamical complexity enables cutting-edge applications related to secure chaos \cite{Shore_99, Uchida_15} and quantum \cite{QKD} communications, rf over fiber communications \cite{Yao_09, Liu_11, Sooudi_15}, and optical sensing \cite{Chow_09}.

The periodic modulation of either the current of the slave laser, or the injection beam, provides an additional degree of freedom that introduces new interesting dynamics in OILs \cite{Simpson_99, Kovanis_01, Kovanis_02, Lin_09}, including Arnold-type locking with additional regions of stable frequency locking and generation of devil's staircases \cite{Lingnau_20}. These features suggest a great potential for the utilization of semiconductor lasers with modulated optical injection as simple signal processing units \cite{Desmet_20} and has driven a recently increasing research interest related to the dynamics of frequency combs injected to OILs \cite{Shortiss_19, Doumbia_20a, Doumbia_20b, AlMulla_20, Quirce_20, Wu_21}.  

The key issue in the dynamical response of such periodically stimulated limit-cycle oscillators is their ability for synchronization between the internal characteristic frequency of the limit cycle and the external driving frequency, in order to produce a phase-locked periodic output \cite{Adler_73, Glass_82, Bak_83, Glass_94}. The synchronization dynamics have global characteristics that have been met in a large variety of physical or technological systems ranging from biological systems \cite{Glass_84, Winfree, Pikovsky, Izhikevich} to electronic circuits \cite{Adler_73}. A stable limit cycle is uniquely characterized in terms of its synchronization properties by the foliation of its basin of attraction as partitioned by its corresponding \textit{isochrons} that determine the \textit{phase response} of the system and allows for the study of synchronization dynamics in terms of one-dimensional circle maps \cite{Winfree, Guckenheimer, Izhikevich}. Although, these concepts have been widely used and proved extremely useful in biological oscillators, their utilization in optical oscillators has not been explored yet. 

In this work, we consider and numerically calculate the phase space structure of the isochrons for limit cycles of an OIL and obtain their corresponding phase response, for the first time, to the best of our knowledge. Moreover, we utilize them in order to study the synchronization dynamics of OIL subjected to periodic modulation of the injected beam in terms of a simple circle map capable of governing the dynamics of the original system and predicting conditions for periodic output under a periodic modulation in the form of a frequency comb.

The fundamental model describing the dynamics of the normalized complex electric field $Y$ and the normalized excess carrier density $Z$ in an Optically Injected Laser is 
\begin{equation}
    \renewcommand{\arraystretch}{2}
    \begin{array}{l}\dfrac{dY}{dt}=(1+i\alpha)YZ-i\Omega Y+\eta\\
    T\dfrac{dZ}{dt}=P-Z-(1+2Z)|Y|^{2} \label{model_1}
    \end{array}
\end{equation}
where $\eta\equiv\sqrt{\dfrac{\tau_{s}G_{N}}{2}}\tau_{p}\kappa E_{in}$ is the normalized injection rate, $\Omega\equiv\nu\tau_{p}$ is the normalized detuning between the frequency of the master laser and the frequency of the free-running slave laser, $\alpha$ is the linewidth enhancement factor, $T$ is the ratio of carrier to photon lifetimes, and $P$ is the normalized excess electrical pumping rate of the slave laser. The rate equation model (\ref{model_1}) can be derived from first principles \cite{Tartwijk_95, Erneux_book} and has been systematically checked against experimental data with an unprecedented agreement \cite{Lenstra_02, Simpson_03, Lindberg_05}. Moreover, the complex dynamical features supported by the model \cite{Phys_Rep} have been confirmed by several experimental studies \cite{Mogensen_85, Simpson_93, Simpson_95, Kovanis_95, Simpson_97}. The model can also be expressed either in terms of the amplitude $R$ and phase $\psi$ ($Y=Re^{i\psi}$) or the real $x$ and imaginary $y$ parts ($Y=x+iy$), of the complex electric field, with the latter given as 
\begin{equation}
    \renewcommand{\arraystretch}{2}
    \begin{array}{l}\dfrac{dx}{dt}=(x-\alpha y)Z+\Omega y+\eta\\
    \dfrac{dy}{dt}=(y+\alpha x)Z-\Omega x \label{model_2}\\
    T\dfrac{dZ}{dt}=P-Z-(1+2Z)(x^2+y^2)
    \end{array}
\end{equation}

This system is well known for its dynamical complexity, including a rich set of qualitatively different dynamical features, including stable and unstable steady states (fixed points), self-sustained oscillations (limit cycles) as well as self-modulated quasiperiodic (torus) and chaotic (strange attractors) oscillations, with all of them related through a web of bifurcations \cite{Phys_Rep}. Among these features, in this work, we are mostly interested in conditions under which the system supports a self-sustained oscillation corresponding to a stable limit cycle. A basic bifurcation analysis is needed in order to identify the relevant parameter ranges in terms of injection rate $\eta$ and detuning $\Omega$ as well as other co-existing dynamical objects. The latter, being unstable, cannot be observed as operating states of the system; however, they have a crucial role in determining the phase space structure of the isochrons and the phase response of the stable limit cycle of interest, as will be shown in the following sections and in the Supplemental Material.

For the case where the system exhibits a stable limit cycle, any initial condition within its basin of attraction asymptotically evolves to the limit cycle. After a transient time interval, all solutions coincide with the limit cycle oscillatory solution, but with a different relative phase of oscillation, in general. The locus of the initial conditions, within the basin of attraction of the limit cycle, that have the same asymptotic phase is known as an \textit{isochron} \cite{Winfree, Guckenheimer, Izhikevich}. In fact, isochrons partition the basin of attraction of a limit cycle according to the most interesting quantity characterizing each initial condition, namely its asymptotic phase. The complement of the basin of attraction of the limit cycle with respect to the whole phase space is called the \textit{phaseless set}. Initial conditions within the phaseless set cannot be characterized by an asymptotic phase value, since they do not evolve to the stable limit cycle. Phaseless sets consist of coexisting fixed points and limit cycles along with their stable manifolds and their basins of attraction, the boundaries of which crucially determine the structure of the isochrons of the limit cycle of interest in the phase space of the system. By introducing these two notions an invariant foliation of the $n$-dimensional phase space is constructed, consisting of (n-1)-dimensional hypersurfaces, i.e. the isochrons.

The isochrons of a system cannot be computed analytically in general. Their numerical computation is a challenging problem, especially for non planar dynamical systems. Most numerical techniques are based either on backward integration for a large number of initial conditions \cite{Winfree, Izhikevich}, or numerical continuation \cite{Osinga_10}. However, backward integration methods suffer from instabilities, especially for strongly stable limit cycles for which backward integration results in strongly diverging reverse orbits, whereas the application of numerical continuation methods presents difficulties in systems with dimension higher than two. In this work, we utilize a qualitatively different alternative method, based on the Koopman operator formalism and the computation of Fourier averages evaluated along trajectories \cite{Mauroy_12}, which is more appropriate for our three-dimensional system.

\begin{figure}  
\begin{center}
    \subfigimg[width=0.70\columnwidth]{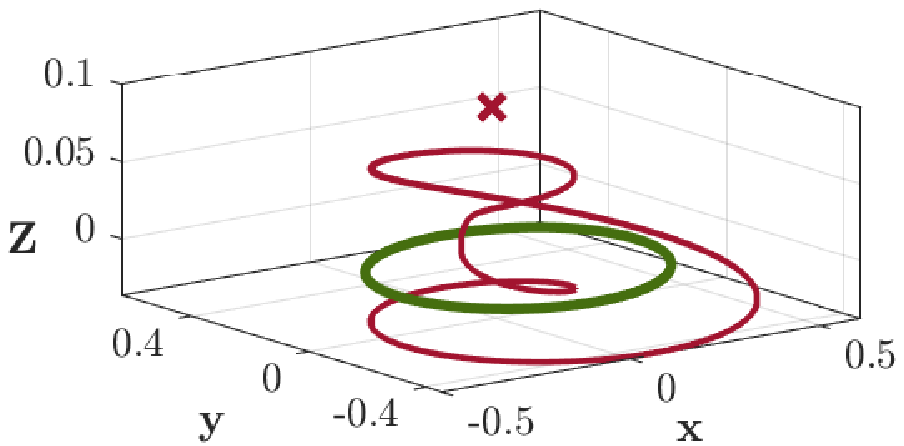}[1\baselineskip]{lc02}\\
    \subfigimg[width=0.70\columnwidth]{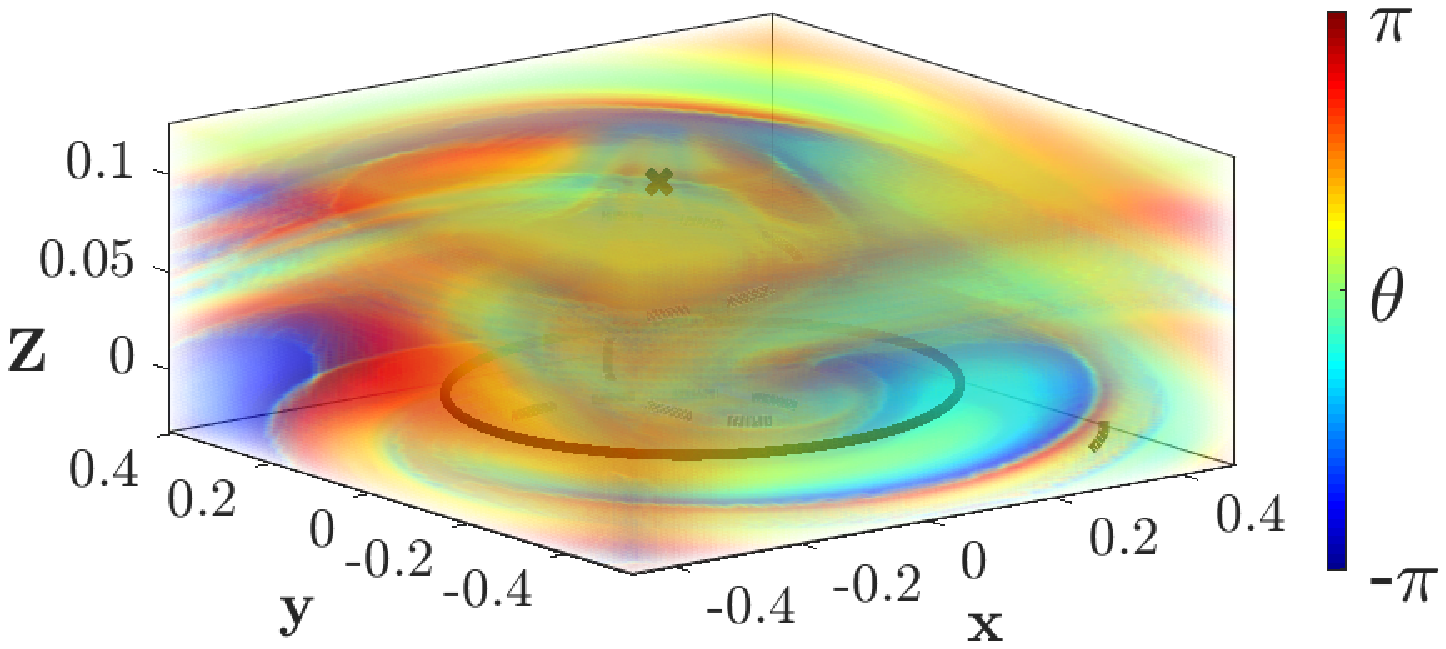}[1\baselineskip]{lc2}\\
    \subfigimg[width=0.70\columnwidth]{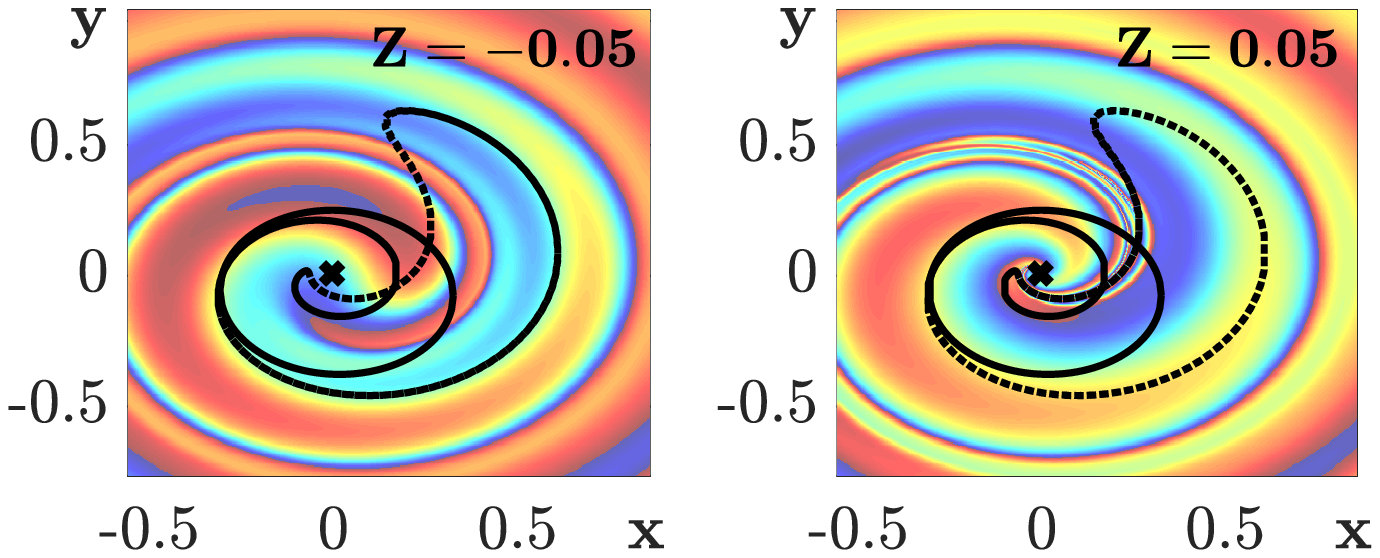}[1\baselineskip]{lc2_sections_z}\\
    \hspace{-1.5em}\subfigimg[width=0.70\columnwidth]{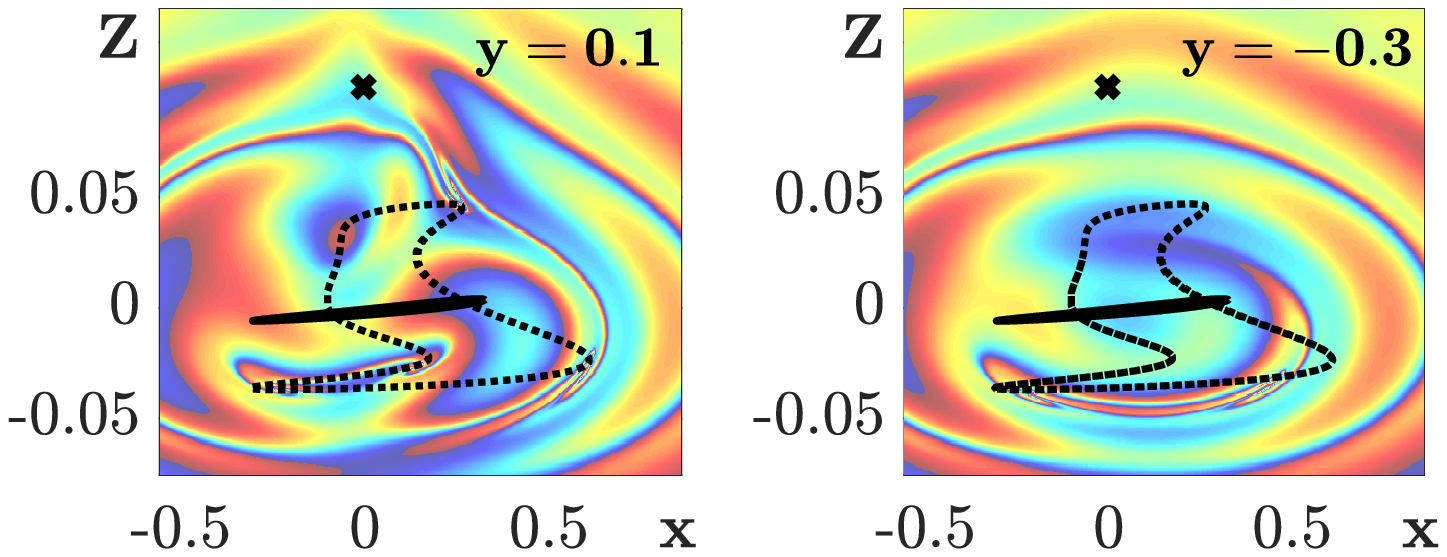}[1\baselineskip]{lc2_sections_y}
    \caption{(a) Phase space of the system (\ref{model_2}) for $\Omega=0.06$, $\eta=0.0025$. x-points denote fixed points, green (thick) and red (light) curves denote stable and unstable limit cycles, respectively. (b) Isochron foliation of the 3D stable manifold of the LC. (c) Sections $\{Z=-0.05\}$, $\{Z=0.05\}$, $\{y=0.1\}$, $\{y=-0.3\}$ of (b). } 
    \label{2nd}
\end{center}
\end{figure}

The structure of the isochrons in the phase space for each stable limit cycle along with its dependence on the phaseless set associated with coexisting stable dynamical objects is systematically shown in the Supplemental Material. For the case shown here in Fig. \ref{2nd} $(\eta, \Omega)=(0.0025, 0.06)$ the phaseless set includes, in addition to the one-dimensional stable manifold of the saddle, a coexisting unstable limit cycle, which significantly complicates the form of the isochrons. 

The complex structure of the isochrons is of paramount practical importance, since it underlies the dynamical response of the system under perturbations due to either external modulation or coupling with an other oscillating system, and determines its synchronization properties. For a system periodically evolving along a stable limit cycle, any stimulus shifts the system from one point of the phase space to a new point having a different asymptotic phase, in general. The \textit{Phase Response Curve} (PRC) is defined as the difference between the new $(\theta_{new})$ and the old $(\theta)$ phase
\begin{equation}
    \text{PRC}(\theta)=\theta_{new}-\theta 
\end{equation}
Positive (negative) values of the PRC correspond to phase advances (delays) with respect the periodic oscillation. Taking into account the amplitude $(A)$ of the stimulus the \textit{generalized} PRC is defined as PRC$(A,\theta)$. Depending on the amplitude of the perturbation the PRC can be either continuous (small amplitude) or discontinuous (large amplitude), characterized as Type 1 or Type 0, respectively.

For the case of a periodic sequence of stimuli with period $T_s$, applied on a system with a limit cycle of period $T_0$, if the $n$-th stimulus is applied when the phase is $\theta_n$ with $\theta_n \in [0, T_0)$, the phase at the moment of the next stimulus $(n+1)$ is 
\begin{equation}
    \theta_{n+1}=\left(\theta_n+\text{PRC}(\theta_n)+T_s\right) \mod T_0 \label{Poincare}
\end{equation}

This equation defines a Poincare mapping of the interval $[0, T_0)$ to itself, which allows to determine the evolution of the phase of the system as an ``orbit'' $\{\theta_n\}$ if the phase at the first application of the stimulus $\theta_1$ is known. If the orbit $\{\theta_n\}$ converges to a fixed point or a closed orbit, synchronization or phase-locking with the external perturbation takes place, respectively. The fixed points of the Poincare mapping (\ref{Poincare}) are given by the equation
\begin{equation}
    \text{PRC}(\theta)=T_0-T_s \label{fp}
\end{equation}
showing that synchronization takes place when the stimulated phase shift compensates for the difference of the two periods (detuning). A direct consequence is that the amplitude of the PRC determines the margin for the detuning in order to have synchronization. Moreover, the slope of the PRC determines the stability of the fixed point and the corresponding synchronized state, with the condition for stability expressed as 
\begin{equation}
    -2<\text{PRC}'(\theta)<0 \label{stability}
\end{equation}
with the prime denoting differentiation with respect to $\theta$ \cite{Izhikevich}. The stability condition is of crucial importance for synchronization in realistic configurations where the modulated injection signal may deviate from exact periodicity due to noise.  

In the following, we focus in the case of parameter values $(\eta, \Omega)=(0.0025, 0.06)$ for which the limit cycle and its corresponding isochrons are depicted in Fig. \ref{2nd} and we consider a periodic sequence of stimuli corresponding to a modulation of the injection rate $\eta$ according to Dirac comb (periodic sequence of delta functions) shifting the system in phase space along the direction of the vector $\Delta x=(0.1,0,0)$ as shown in Fig. \ref{section}, where the isochrons of the system are depicted in a $Z=0$ cut of the phase space. The projection of the limit cycle is shown and the point where it intersects the plane $Z=0$ is denoted by $B_0$, whereas the point $B$ denotes the point where the system is shifted after the stimulus. The corresponding phase response curve for such perturbation is depicted in Fig. \ref{prc}. It is worth emphasizing that the form of the PRC is determined by the structure of the isochrons in the phase space. Fixed points of the respective Poincare map (\ref{Poincare}) correspond to points of intersection between the PRC and the horizontal line at the level $T_0-T_s$, with parts of the PRC fulfilling the stability condition (\ref{stability}) denoted with a continuous line. The result of the application of a Dirac comb on the asymptotic phase is shown in Fig. \ref{time_series}. 

The period of the limit cycle, for the aforementioned parameter values, is $T_0\simeq99.5865$. In order to have synchronization, the Dirac comb must have a period $T_s$ close to $T_0$, according to Eqs. (\ref{fp})-(\ref{stability}). For $T_s\simeq87.5264$ the horizontal line $y=T_0-T_s$ intersects the curve PRC$(\theta)$ at two points at $\theta_u\simeq72.4447$ and $\theta_s =90$, corresponding to an unstable and a stable fixed point of the Poincare map, respectively, as shown in Fig. \ref{prc}. In such case, starting from a random initial phase (excluding the unstable fixed point $\theta_u$) the orbit of the Poincare map converges to $\theta_s$ as shown in Fig. \ref{poin_orbit_1} and the cobweb plot in Fig. \ref{phm_1}. By applying the same periodic perturbation with a period $100T_0+T_s$ in order to omit the transient stage of the system evolving towards the limit cycle, the phase converges to the predicted values (Fig. \ref{poin_lc_1}) and synchronization takes place as shown in Fig. \ref{orbit_1}.
 
The importance of the information provided by the phase response with respect to synchronization, becomes obvious if, ignoring the restrictions imposed by the PRC$(\theta)$, we choose $T'_s=140$ for which there is no intersection between the line $y=T_0-T'_s$ as shown in Fig. \ref{prc}. In this case the orbit of the Poincare map is complex as shown in Fig. \ref{poin_orbit_2} and the cobweb plot Fig. \ref{phm_2}, and the application of an external perturbation with period $100T_0+T_s'$ results in an irregular aperiodic evolution of the system as shown in Figs. \ref{poin_lc_2} and \ref{orbit_2}. This complex evolution can be either quasiperiodic or chaotic depending on whether the PRC is Type 1 (as in this case) or Type 0 \cite{Glass_94}.

\begin{figure}  
\begin{center}
    \subfigimg[width=0.49\columnwidth]{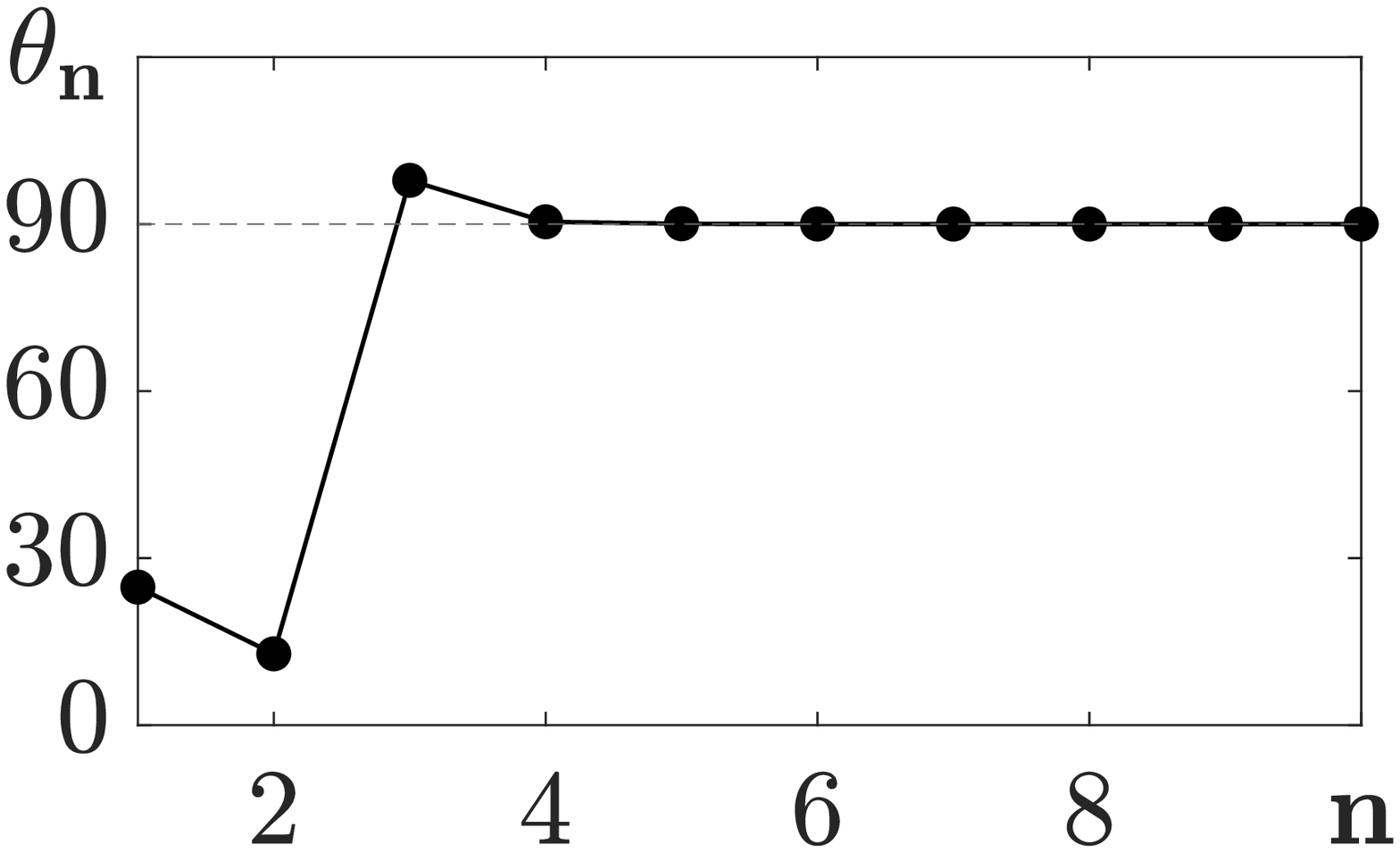}[1\baselineskip]{poin_orbit_1}
    \subfigimg[width=0.49\columnwidth]{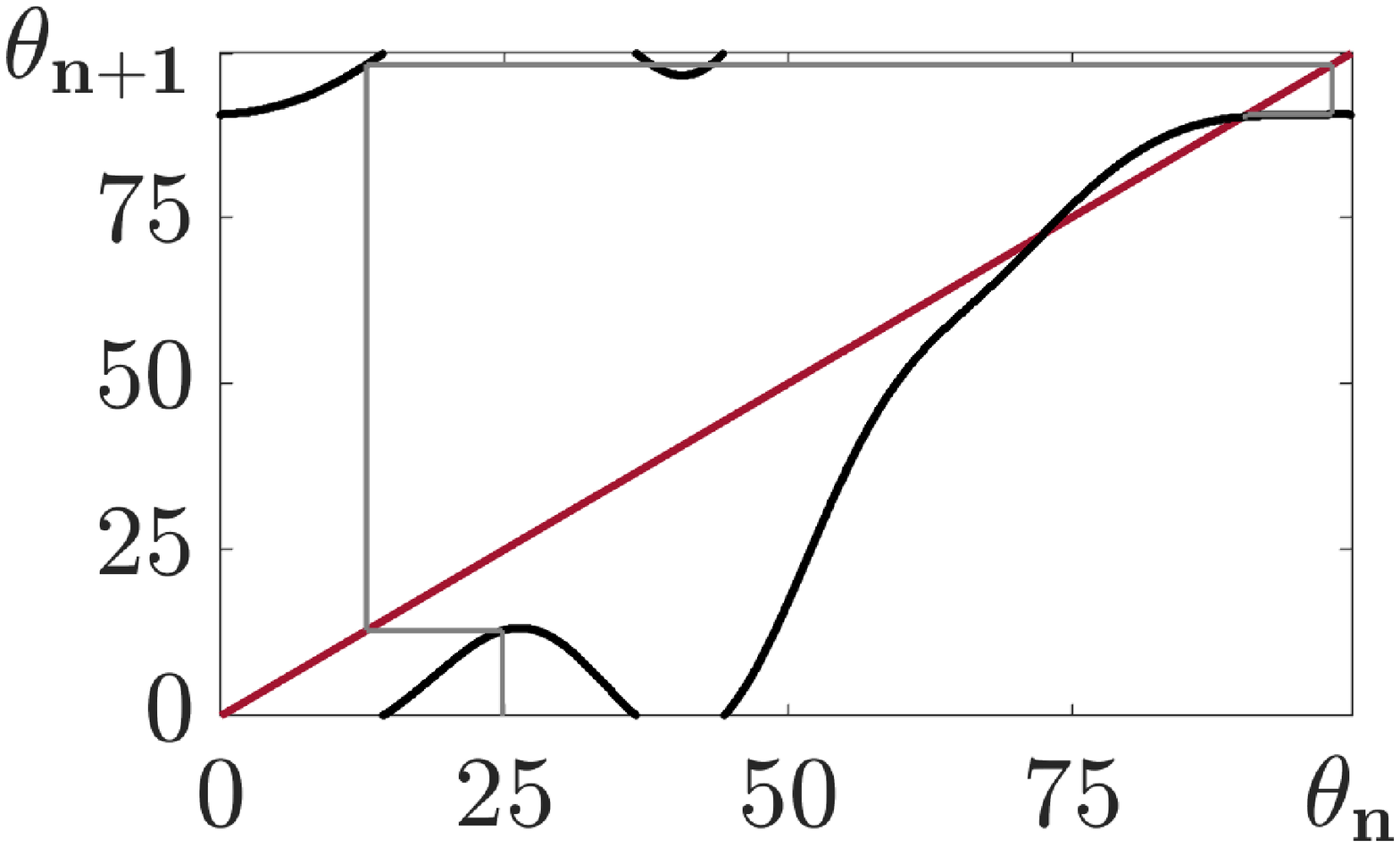}[1\baselineskip]{phm_1}
    \subfigimg[width=0.49\columnwidth]{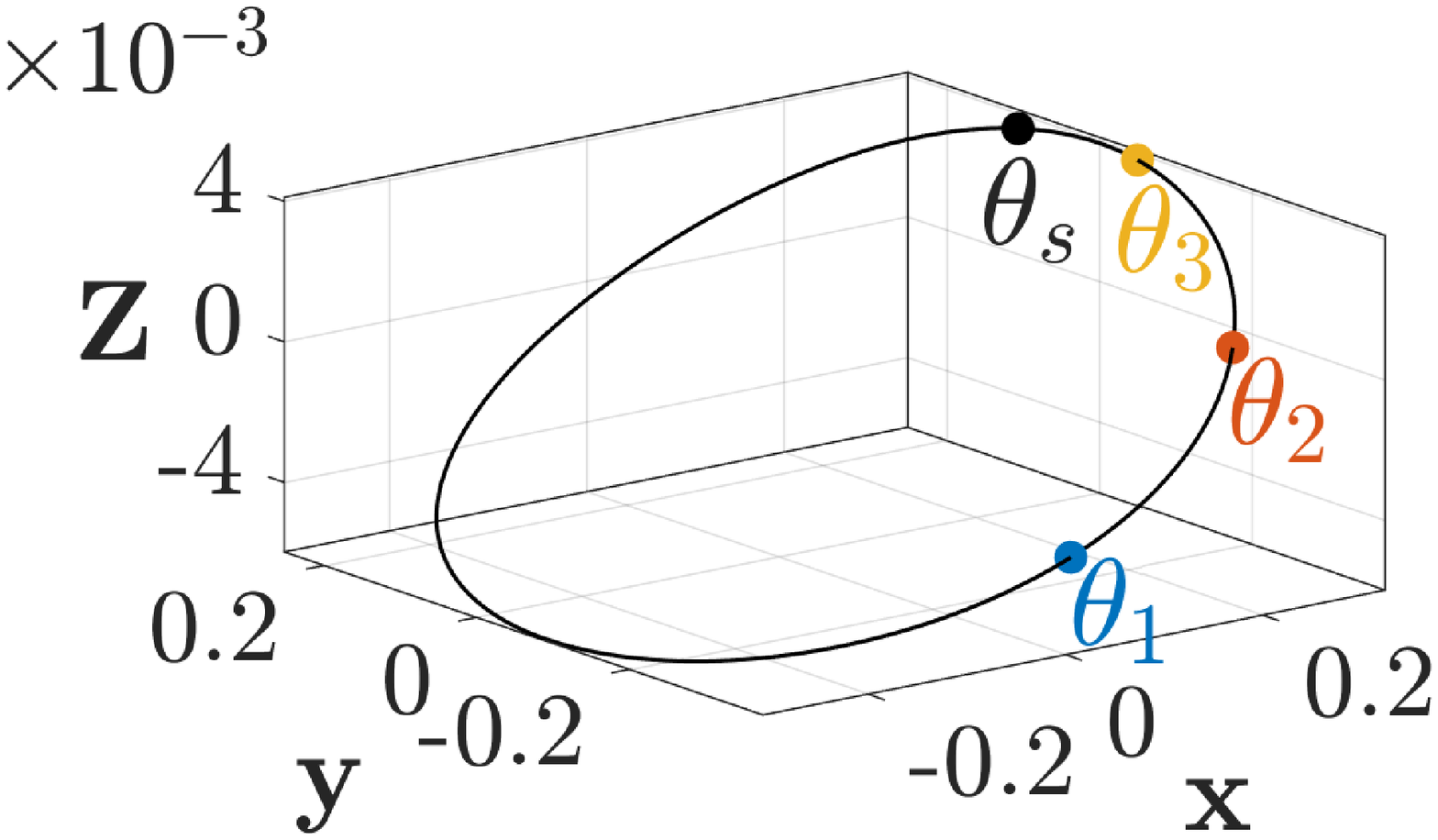}[1\baselineskip]{poin_lc_1}
    \subfigimg[width=0.49\columnwidth]{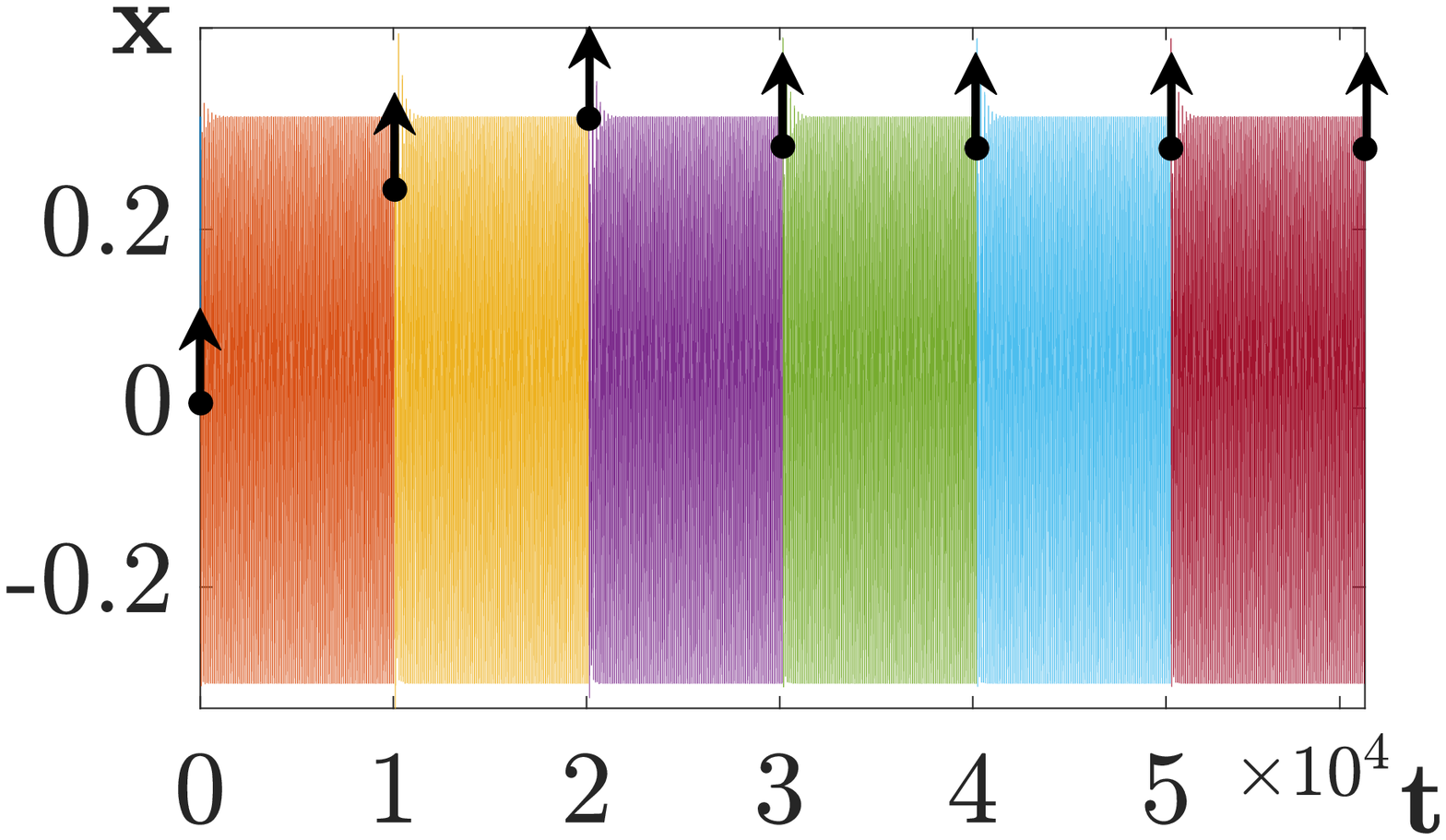}[1\baselineskip]{orbit_1}
    \caption{Synchronization of phase due to the application of a Dirac comb of period $T_s\simeq87.5264$. (a) Convergence of the phase orbit with initial point $\theta_1\simeq24.8219$; (b) Poincare phase map of the orbit shown in (a); (c) Synchronization on the limit cycle; (d) Time series of the perturbed $x$-coordinate as obtained from the original system (\ref{model_2}).} 
    \label{synchronized}
\end{center}
\end{figure}

It is worth emphasizing the spectral differences between a synchronized and an irregular aperiodic evolution of the system. Under the absence of any external perturbation the system evolves periodically (after an initial transient stage) according to its limit cycle. In such case the operating state of the system has a discrete spectrum  with frequency peaks corresponding to harmonics of the fundamental frequency of the limit cycle ($f_0=1/T_0$), due to the nonharmonic form of the limit cycle, as shown in Fig. \ref{psd}. For the synchronized case, corresponding to Fig. \ref{synchronized}, the spectrum is again discrete and forms a finer comb with major peaks corresponding to the harmonics of $f_0$ and significantly high secondary peaks at the harmonics of the frequency of the external perturbation $f_s=1/(kT_0+T_s)$ with $k$ being the number of limit cycle periods ($k=100$ in the aforementioned cases) between the external stimuli which can be utilized for the control of the spacing between the frequencies of the comb, as depicted in Fig. \ref{psd_1}. In contrast to the synchronized state, in the nonsynchronized state the frequency spectrum contains additional spectral components and does not form a clear frequency comb, as shown in Fig. \ref{psd_2}.    

In conclusion, the global structure of the isochrons in the three-dimensional phase space has been considered and numerically calculated for the fundamental oscillator consisted of an Optically Injected Laser for the first time. The drastic dependency of the isochrons on the phaseless sets of the system is shown for characteristic cases of stable limit cycles. Based on the isochrons structure, the phase response of the system has been obtained and the corresponding phase response curves have been calculated. The latter were shown to have a crucial importance for the study of synchronization dynamics of the original three-dimensional dynamical system with modulated injection beam in terms of a reduced one-dimensional circle map. The case of a Dirac comb modulation has been considered for illustration purposes, whereas the method directly applies also to any periodic sequence of pulses with finite duration and arbitrary shape. The study of the dynamics of the circle map was proved capable of defining stable phase-locking conditions for the original system, ensuring a periodic output with a discrete spectrum consisting of equidistant lines in the form of a frequency comb. These results clearly show the importance of the isochrons and the corresponding phase response curves for the study of the complex synchronization dynamics and the capabilities for tunable frequency comb generation in driven photonic oscillators.


\clearpage

\begin{figure}
\begin{center}
    \subfigimg[width=0.44\columnwidth]{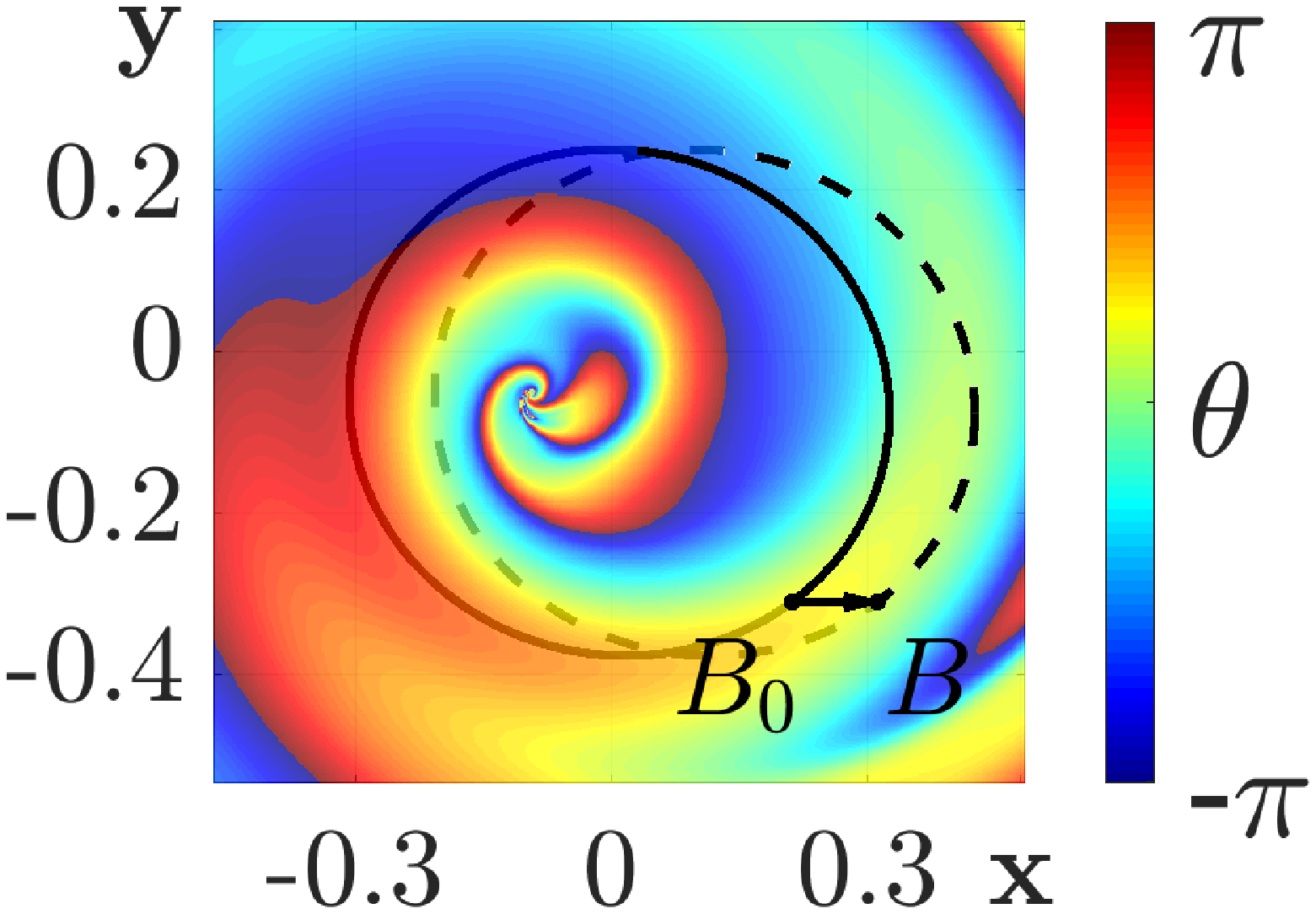}[1\baselineskip]{section}
    \subfigimg[width=0.54\columnwidth]{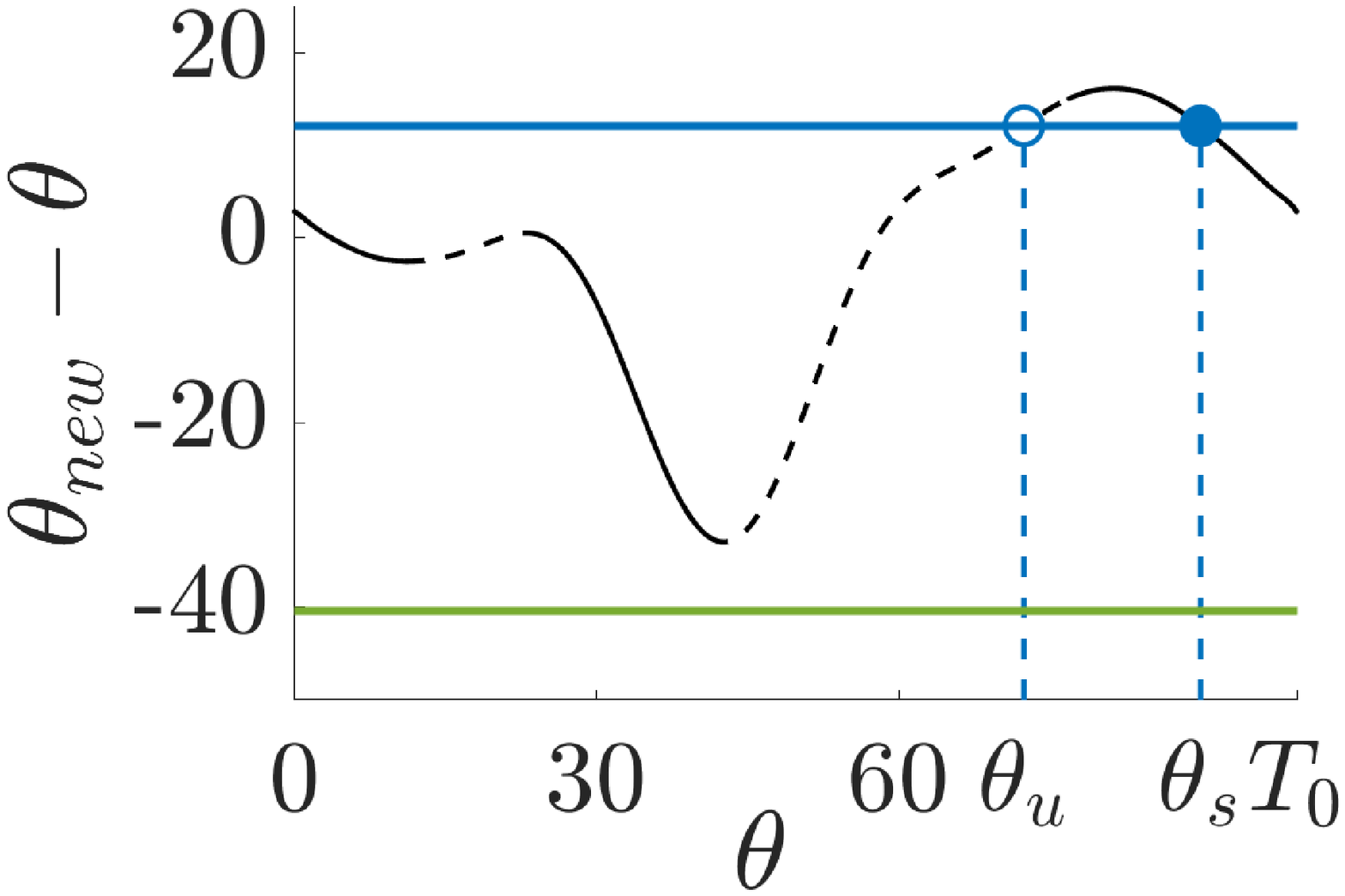}[1\baselineskip]{prc}\\
    \subfigimg[width=1\columnwidth]{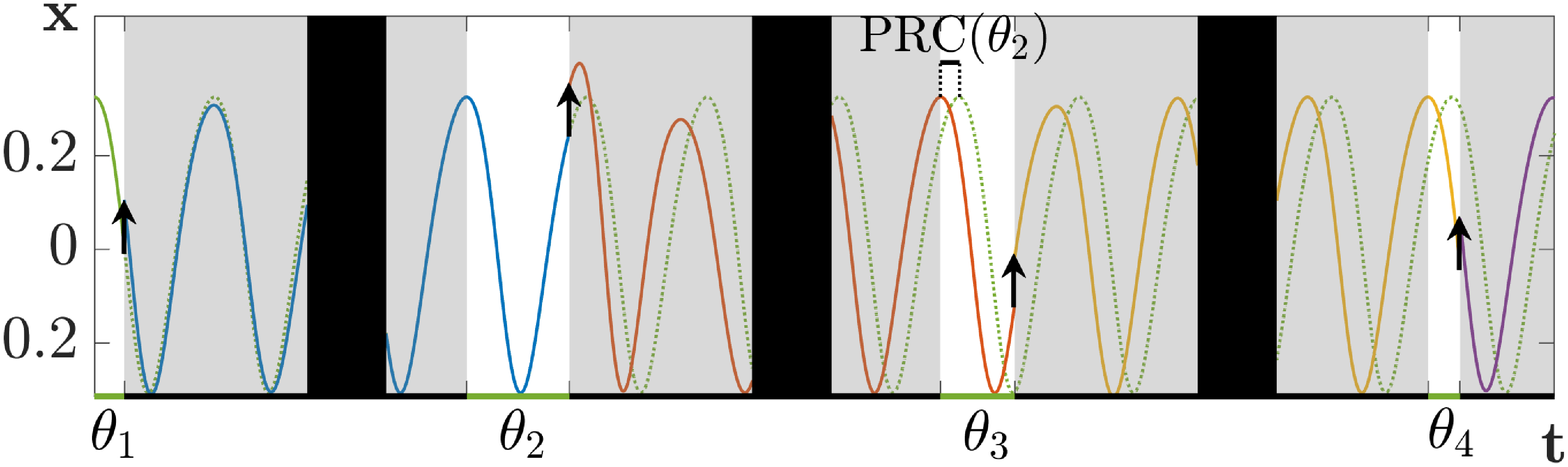}[1\baselineskip]{time_series}
    \caption{ (a) Isochron foliation on $\{Z=0\}$ section of the phase space of (\ref{model_2}) for $\Omega=0.06,\ \eta=0.0025$.  (b) PRC for perturbation $\Delta x=(0.1,0,0).$ The blue line corresponds to $T_0-T_s\simeq12.0601$, while the green line corresponds to $T_0-T'_s\simeq-40.4135$. (c) Time series of perturbed $x$-coordinate. Each solid line corresponds to the evolution of $x$ after every perturbation (denoted by a black arrow), and the green dashed line corresponds to the free-running trajectory. Grey areas represent the first and last part of transient stages, while black areas are the part of the transient stage that is not presented for simplicity reasons.}
\end{center}
\end{figure}
\newpage

\begin{figure}  
\begin{center}
    \subfigimg[width=0.49\columnwidth]{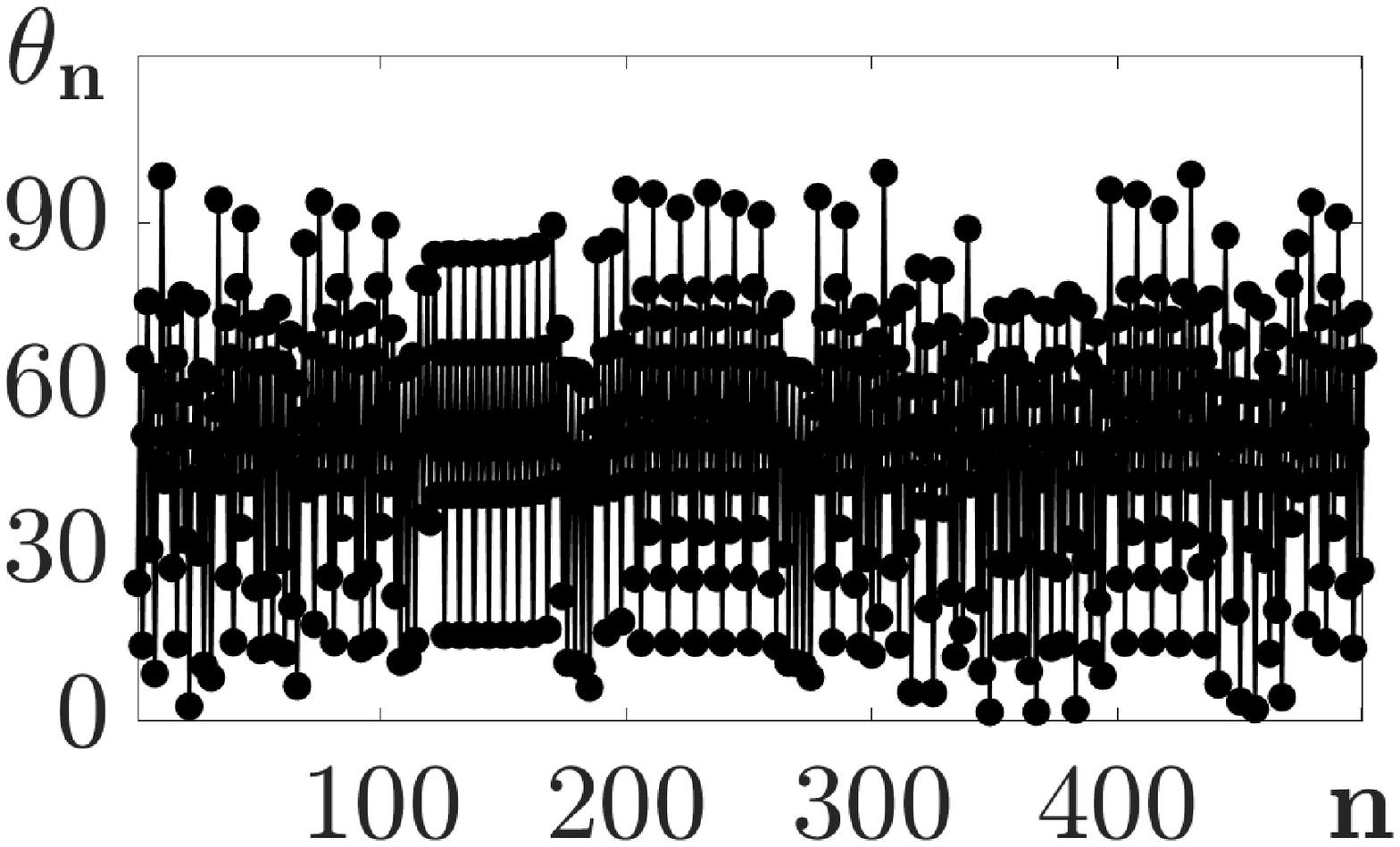}[1\baselineskip]{poin_orbit_2}
    \subfigimg[width=0.49\columnwidth]{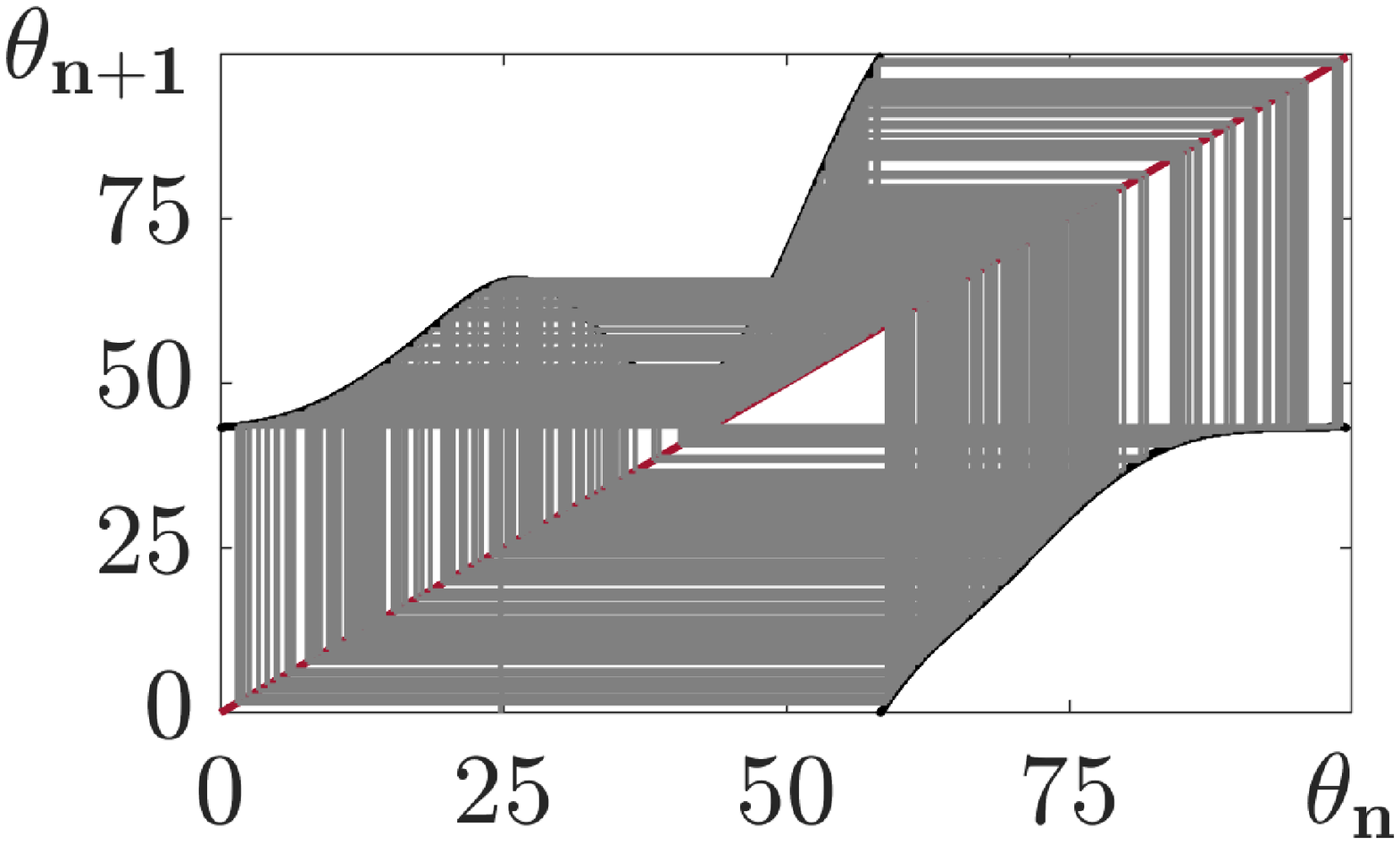}[1\baselineskip]{phm_2}
    \subfigimg[width=0.49\columnwidth]{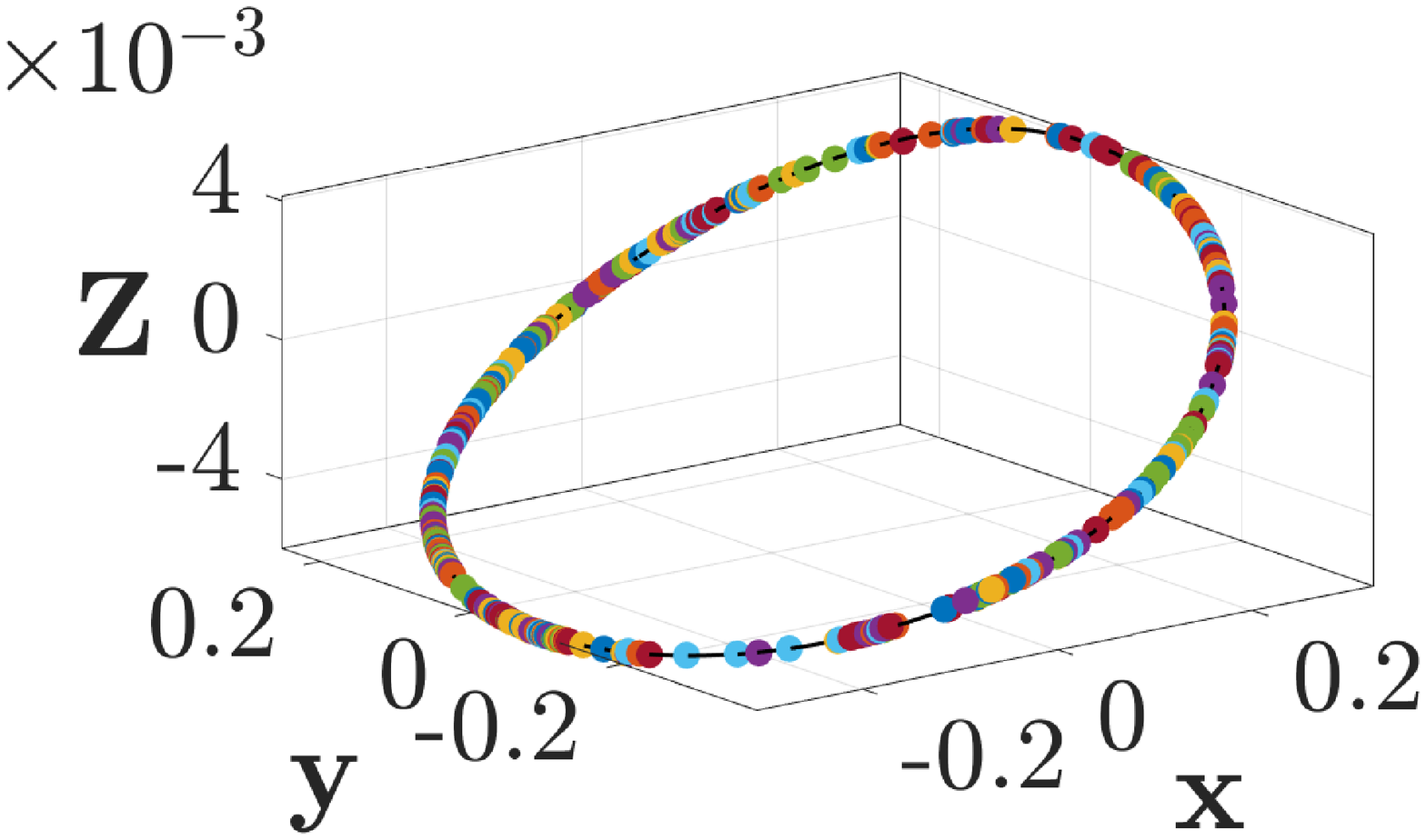}[1\baselineskip]{poin_lc_2}
    \subfigimg[width=0.49\columnwidth]{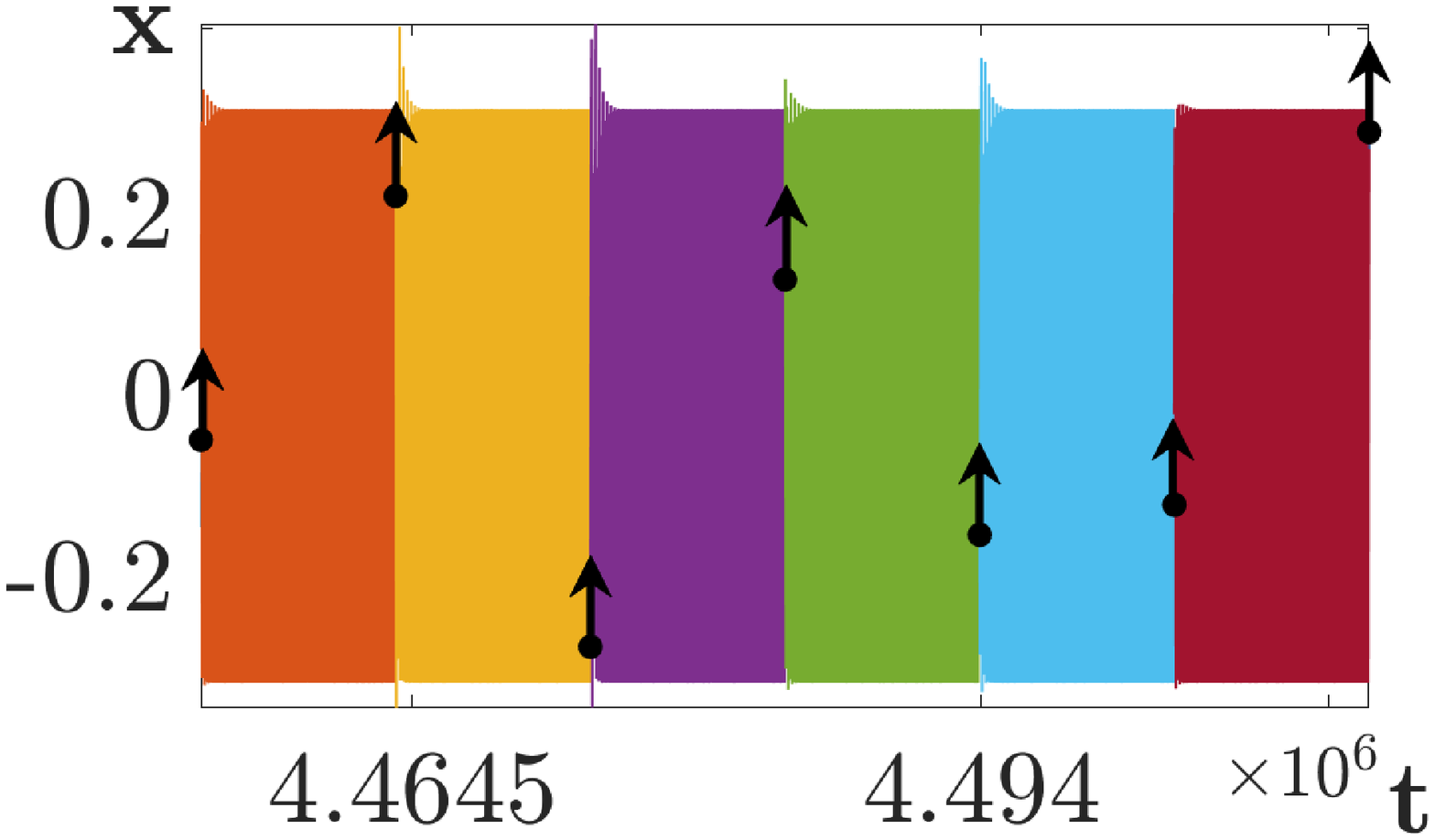}[1\baselineskip]{orbit_2}
    \caption{Complex evolution of phase due to the application of a Dirac comb of period $T'_s=140$. (a) Complex phase orbit with initial point $\theta_1\simeq24.8219$; (b) Poincare phase map of the orbit shown in (a); (c) Phase evolution on the limit cycle; (d) Part of irregular aperiodic time series of the perturbed $x$-coordinate as obtained from the original system (\ref{model_2}).} 
    \label{chaotic}
\end{center}
\end{figure}

\begin{figure}  
\begin{center}
    \subfigimg[width=0.90\columnwidth]{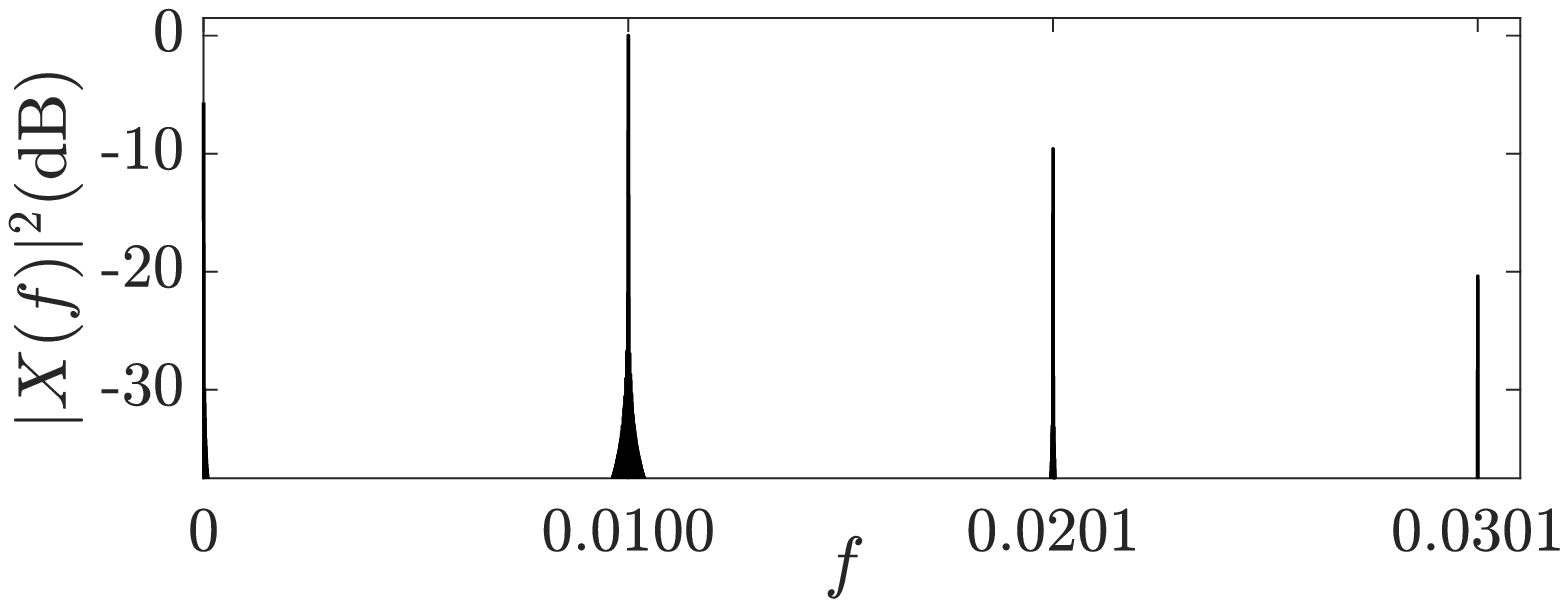}[1\baselineskip]{psd}
    \subfigimg[width=0.90\columnwidth]{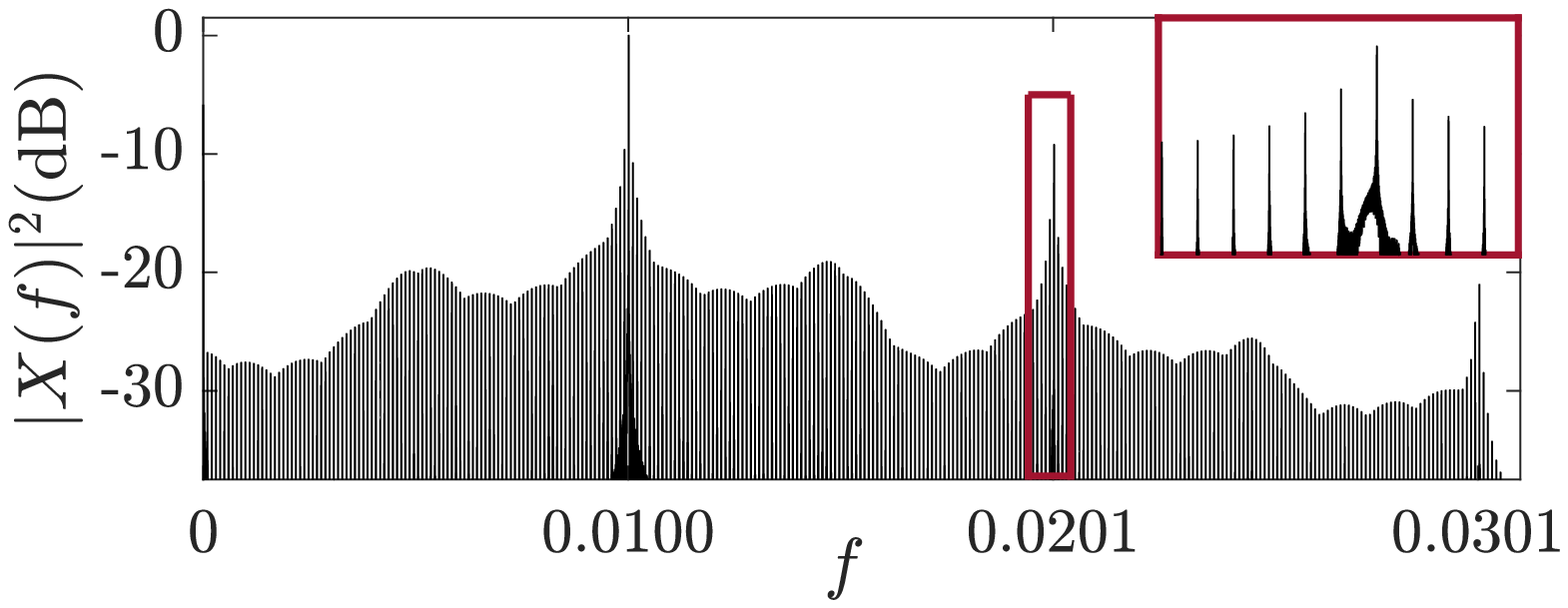}[1\baselineskip]{psd_1}
    \subfigimg[width=0.90\columnwidth]{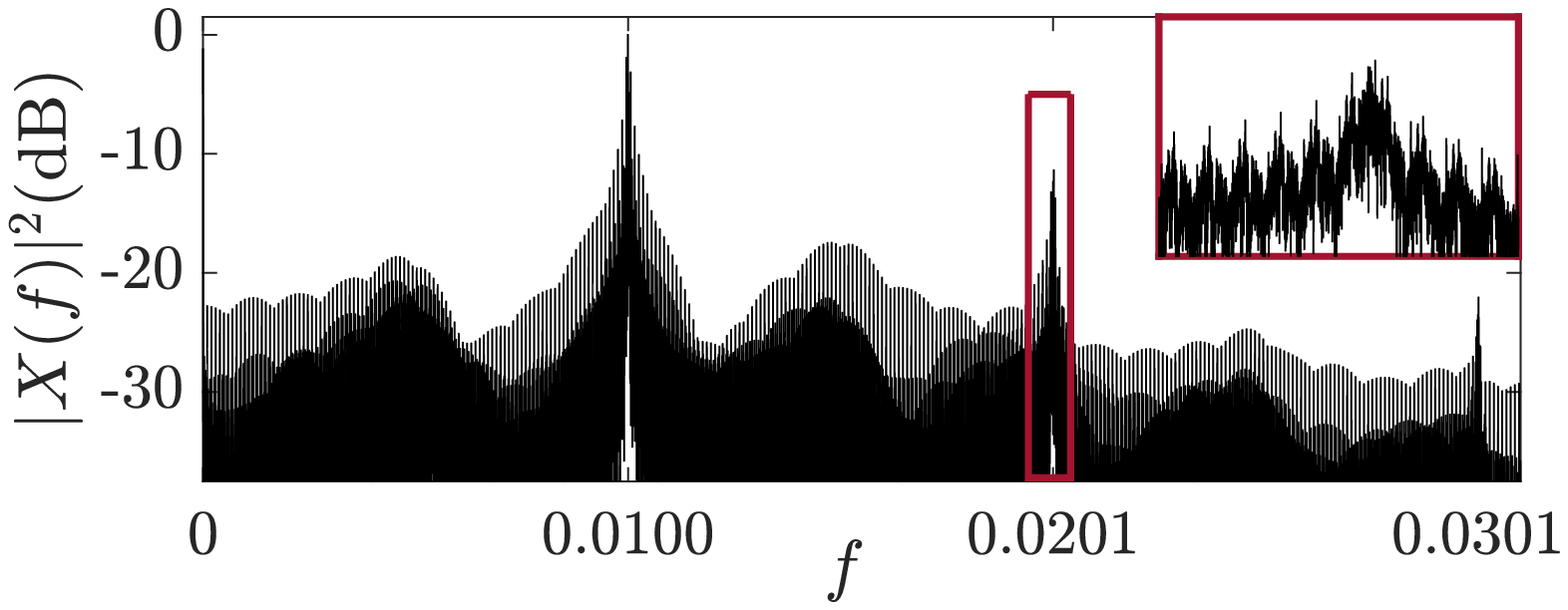}[1\baselineskip]{psd_2}
    \caption{Power spectral density $|X(f)|^2$ for (a) the periodic orbit corresponding to the unperturbed limit cycle, (b) the synchronized time series corresponding to Fig. \ref{orbit_1}, and (c) the irregular aperiodic time series corresponding to Fig. \ref{orbit_2}.  } 
\end{center}
\end{figure}

\newpage
\begin{center} \Large{SUPPLEMENTAL MATERIAL} \end{center}

\section*{Bifurcations}
\begin{figure}[h]
    \begin{center}
         \subfigure{\includegraphics[width=1\columnwidth]{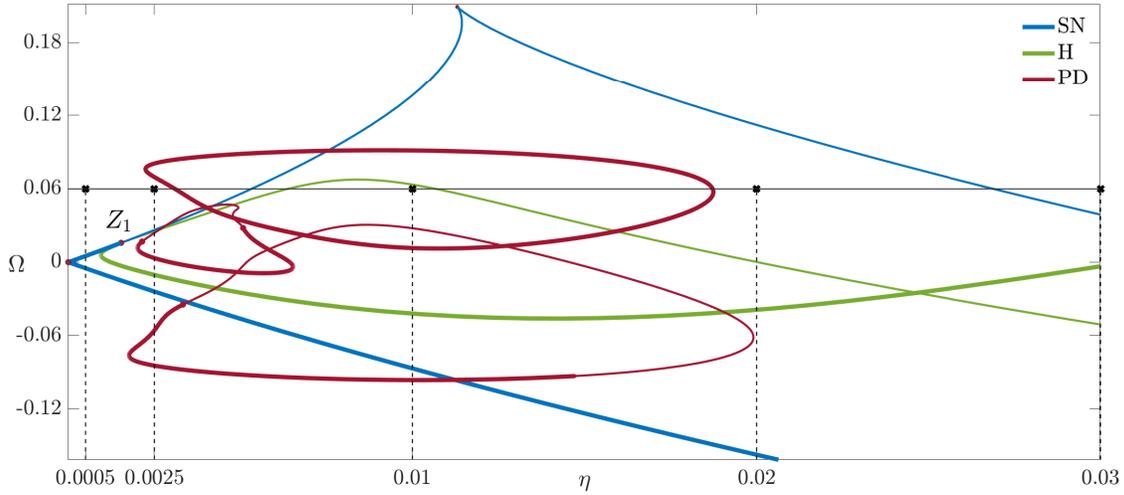}}
         \caption{(Color online) Bifurcation diagram of the $(\Omega, \eta)$ parameter subspace for $\alpha=3$, $T=125$, $P=0.1$. Solid lines represent bifurcations of stationary points (blue: Saddle-Node, green: Hopf) and limit cycles (red: Period-Doubling). Supercritical bifurcations are illustrated by bold lines, while subcritical bifurcations by thinner ones. SN and H curves become tangent at $Z_1$, where a Zero-Hopf bifurcation occurs and the stability of the bifurcation changes. Black points correspond to typical parameter values used in the following figures.}
    \label{fig:bifurcation diagram}
    \end{center}
\end{figure}

The basic bifurcations of the system in the parameter space $(\eta, \Omega)$, are depicted in Fig. \ref{fig:bifurcation diagram}, for $\alpha=3$, $T=125$ and $P=0.1$, where Saddle-Node (SN) bifurcations of fixed points as well as Hopf (H) bifurcations giving rise to Limit Cycles (LC) along with Period Doubling (PD) bifurcations are shown, as obtained with the utilization of the numerical continuation toolbox MatCont [1].  The parameter space is dissected for a fixed value of the detuning ($\Omega=0.06$) and varying values of the injection rate $\eta$, and specific values are given for the points of section between the horizontal line $\Omega=0.06$ and the bifurcation curves in Fig. \ref{fig:bifurcation diagram}. The corresponding dynamical objects of interest, namely limit cycles and fixed points are depicted in the three-dimensional phase space as in Figs. \ref{lc01}-\ref{lc05}. For the fixed value of $\Omega=0.06$, starting from a small value of the injection rate $(\eta=0.0005)$, the system has a saddle-focus fixed point and a stable limit cycle. As the injection rate is increased $(\eta=0.0025)$ an unstable limit cycle also appears, whereas further increasing $(\eta=0.01)$ beyond the period doubling bifurcation value $(\eta=0.003092)$ renders the limit cycle unstable. For $\Omega$ and $\eta$ values located within the region bounded by the SN bifurcation curves in Fig. \ref{fig:bifurcation diagram}, there exist either two saddle-foci and one unstable focus-node $(\eta\in(0.00533,0.00639)\cup(0.01065,0.02688),\ \Omega=0.06)$ or three saddle-foci $(\eta\in(0.00639,0.01065),\ \Omega=0.06)$; the system has a single fixed point corresponding to a stable focus-node for parameter values on the right hand side of the H curve. 

\section*{Isochron structures}
\begin{figure}[h]  
\begin{center}
     \subfigimg[width=0.49\columnwidth]{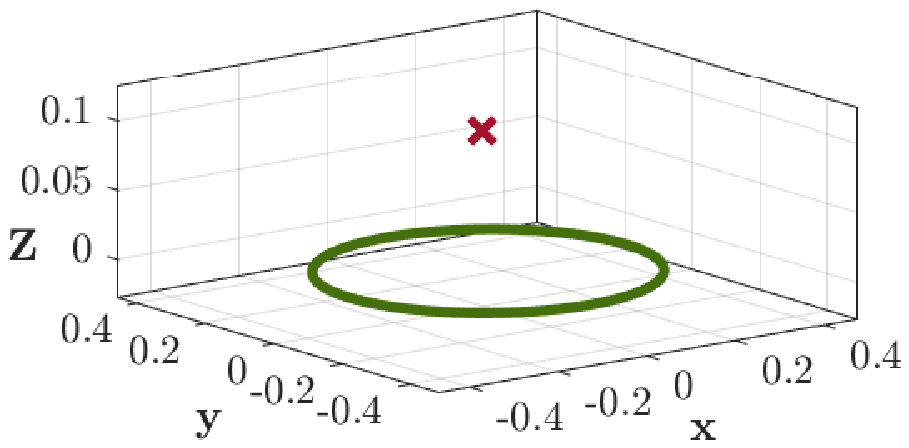}[1\baselineskip]{lc01}
     \subfigimg[width=0.49\columnwidth]{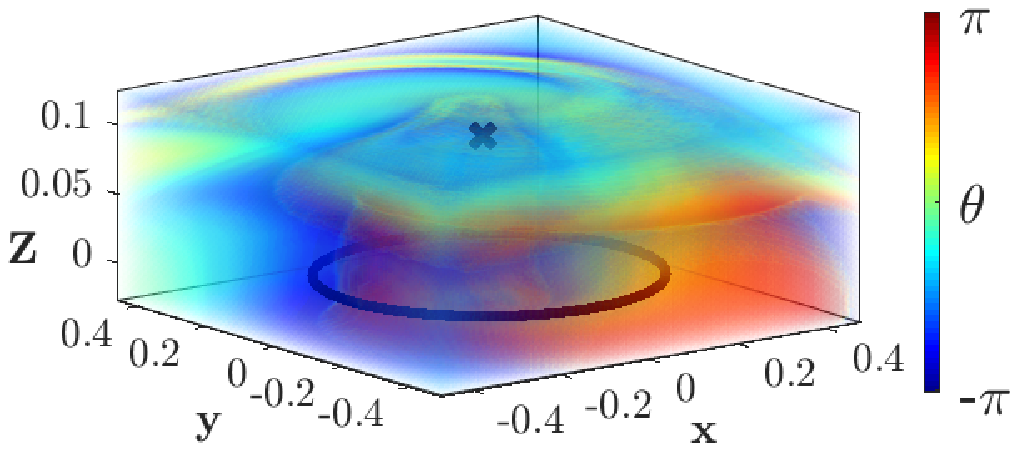}[1\baselineskip]{lc1}
    \subfigimg[width=0.48\columnwidth]{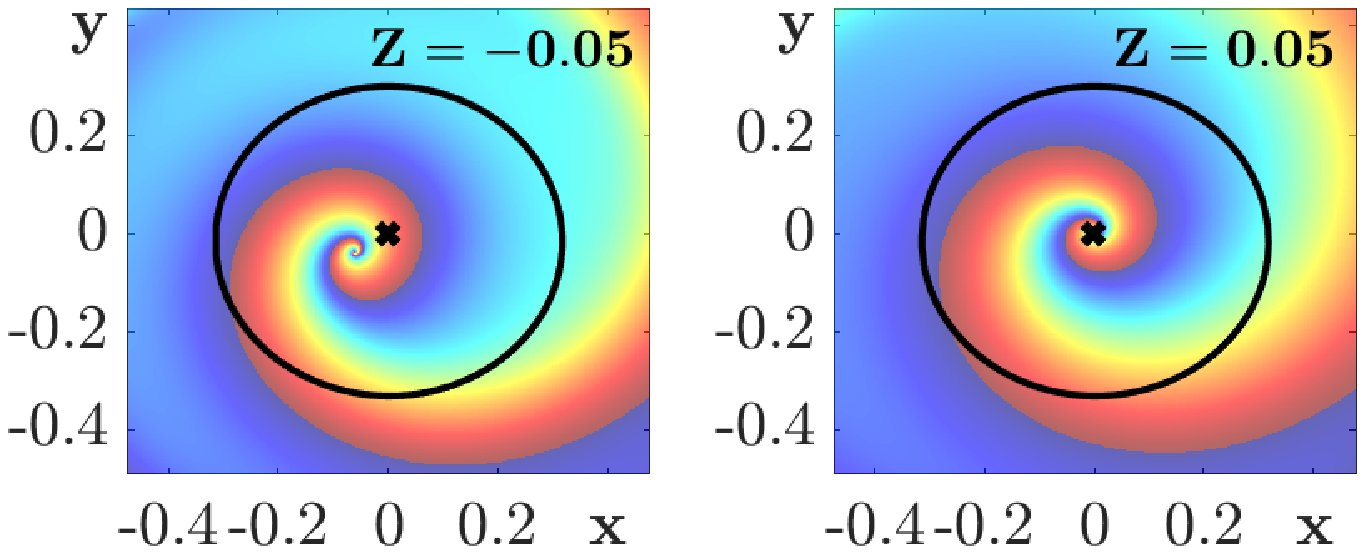}[1\baselineskip]{lc1_sections_z}
    \subfigimg[width=0.48\columnwidth]{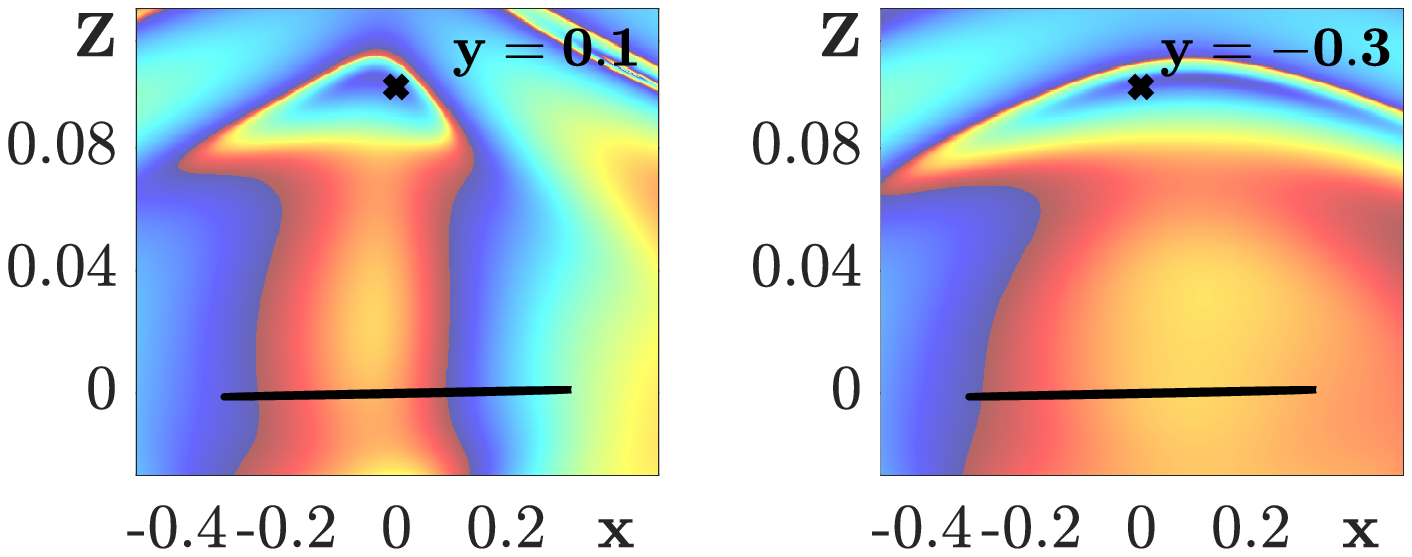}[1\baselineskip]{lc1_sections_y}
     \caption{(a) Phase space of the system for $\Omega=0.06$, $\eta=0.0005$. x-points denote fixed points, green (thick) and red (light) curves denote stable and unstable limit cycles, respectively. (b) Isochron foliation of the 3D stable manifold of the LC. (c) Sections $\{Z=-0.05\}$, $\{Z=0.05\}$, $\{y=0.1\}$, $\{y=-0.3\}$ of (b). } 
     \label{1st}
\end{center}
\end{figure}

\begin{figure}  
\begin{center}
    \subfigimg[width=0.49\columnwidth]{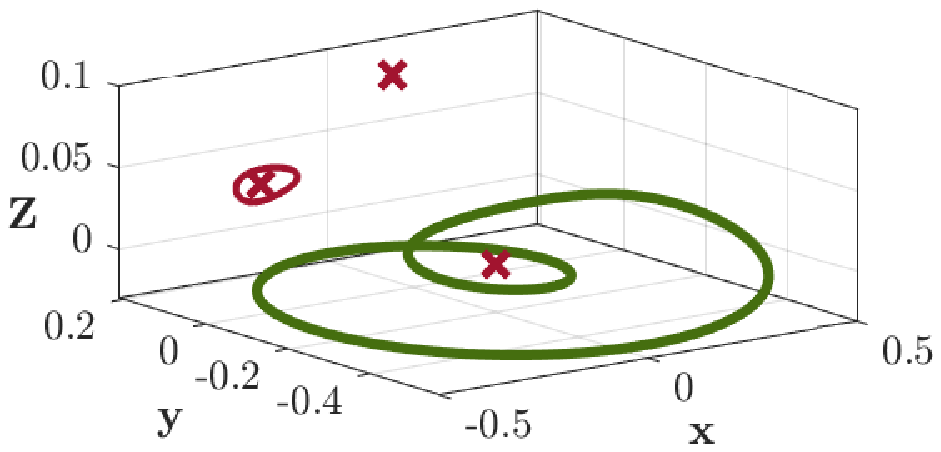}[1\baselineskip]{lc03}
    \subfigimg[width=0.49\columnwidth]{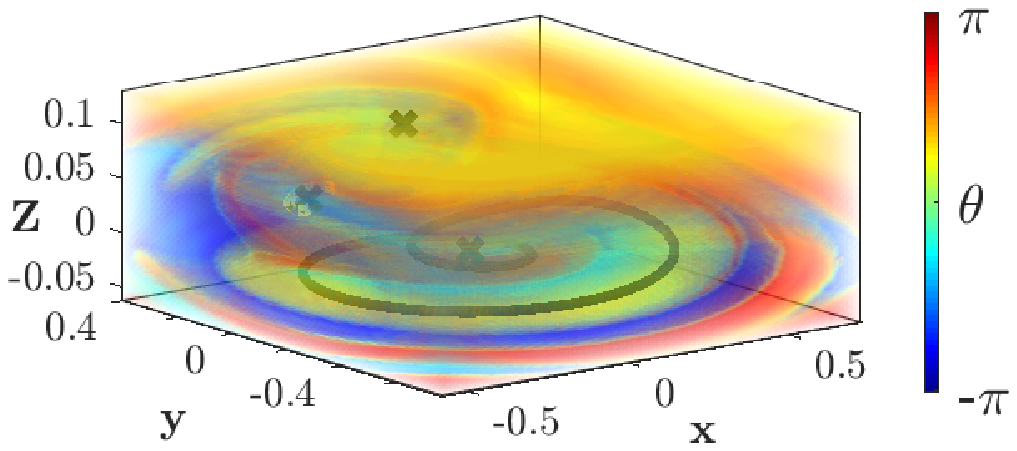}[1\baselineskip]{lc3}
    \subfigimg[width=0.48\columnwidth]{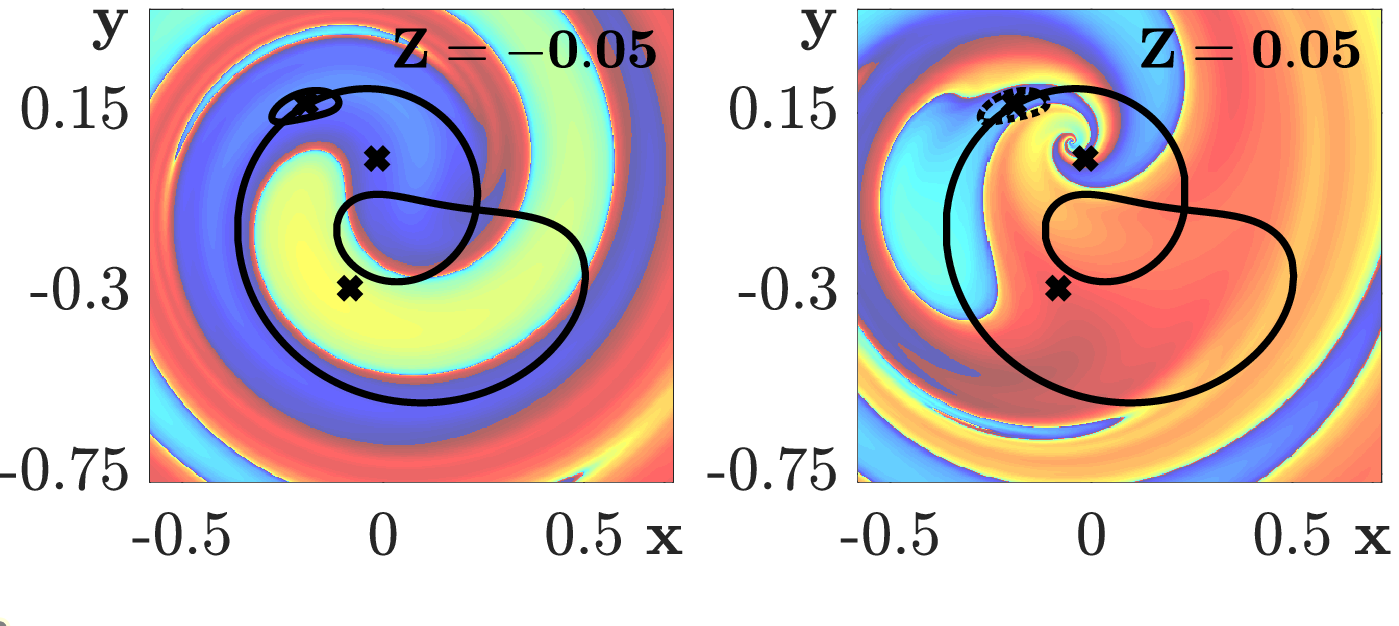}[1\baselineskip]{lc3_sections_z}
    \subfigimg[width=0.48\columnwidth]{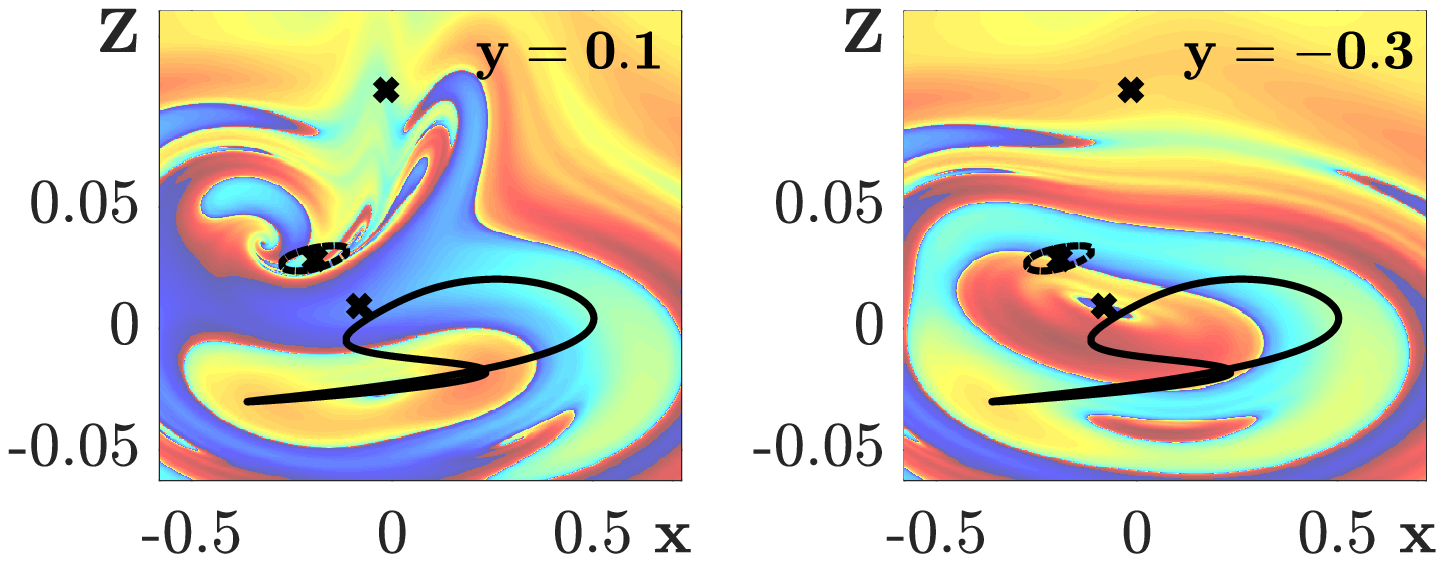}[1\baselineskip]{lc3_sections_y}
    \caption{(a) Phase space of the system for $\Omega=0.06$, $\eta=0.01$. x-points denote fixed points, green (thick) and red (light) curves denote stable and unstable limit cycles, respectively. (b) Isochron foliation of the 3D stable manifold of the LC. (c) Sections $\{Z=-0.05\}$, $\{Z=0.05\}$, $\{y=0.1\}$, $\{y=-0.3\}$ of (b). } 
    \label{3rd}
\end{center}
\end{figure}

\begin{figure}  
\begin{center}
    \subfigimg[width=0.49\columnwidth]{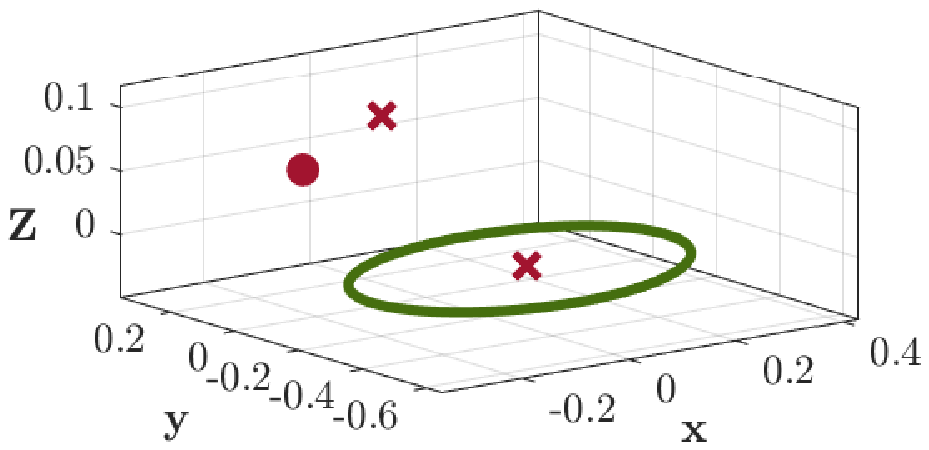}[1\baselineskip]{lc04}
    \subfigimg[width=0.49\columnwidth]{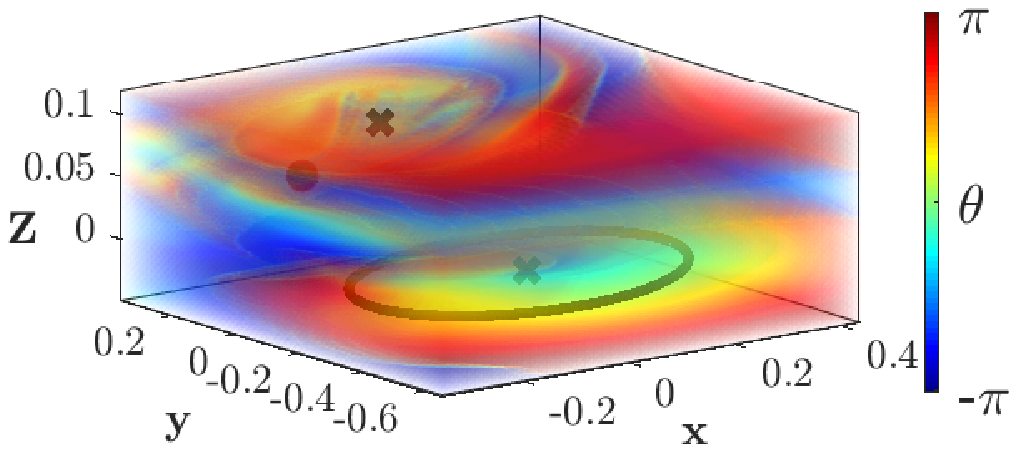}[1\baselineskip]{lc4}
    \subfigimg[width=0.48\columnwidth]{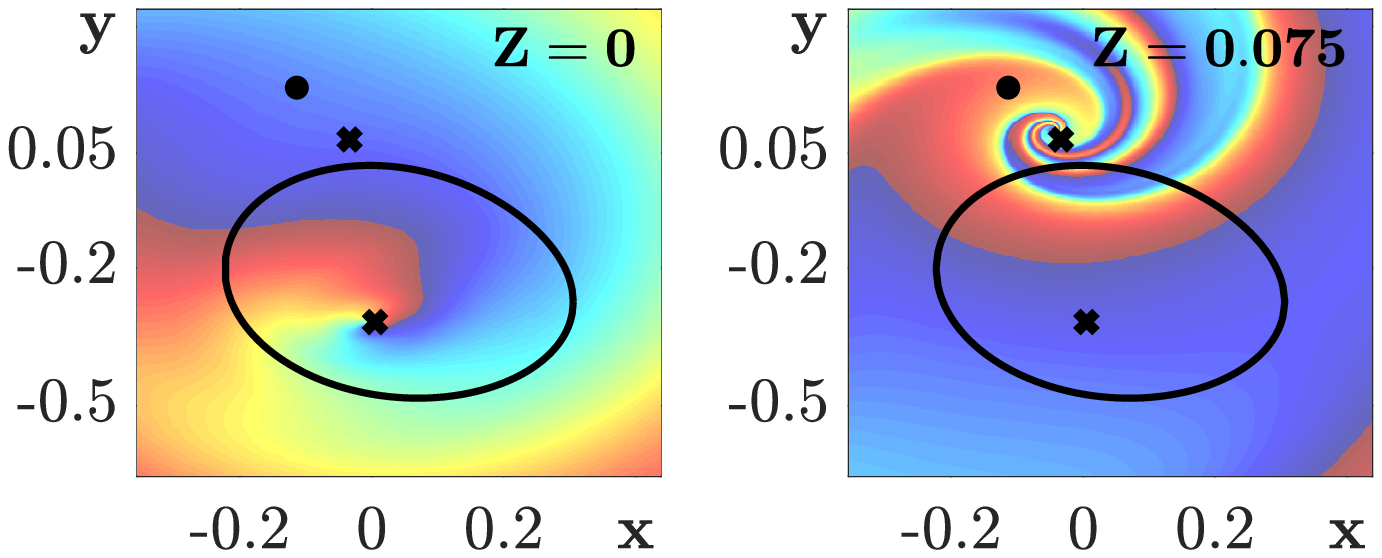}[1\baselineskip]{lc4_sections_z}
    \subfigimg[width=0.48\columnwidth]{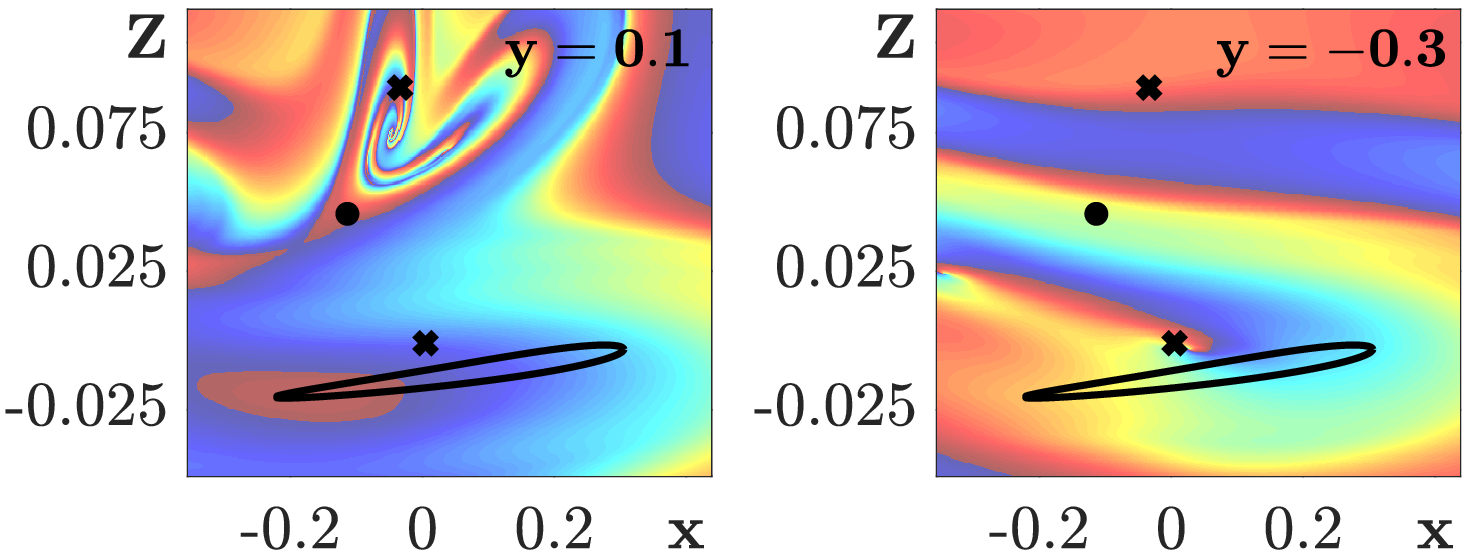}[1\baselineskip]{lc4_sections_y}
    \caption{(a) Phase space of the system for $\Omega=0.06$, $\eta=0.02$. x-points denote fixed points, green (thick) and red (light) curves denote stable and unstable limit cycles, respectively. (b) Isochron foliation of the 3D stable manifold of the LC. (c) Sections $\{Z=0\}$, $\{Z=0.075\}$, $\{y=0.1\}$, $\{y=-0.3\}$ of (b). } 
    \label{4th}
\end{center}
\end{figure}

\begin{figure}  
\begin{center}
    \subfigimg[width=0.49\columnwidth]{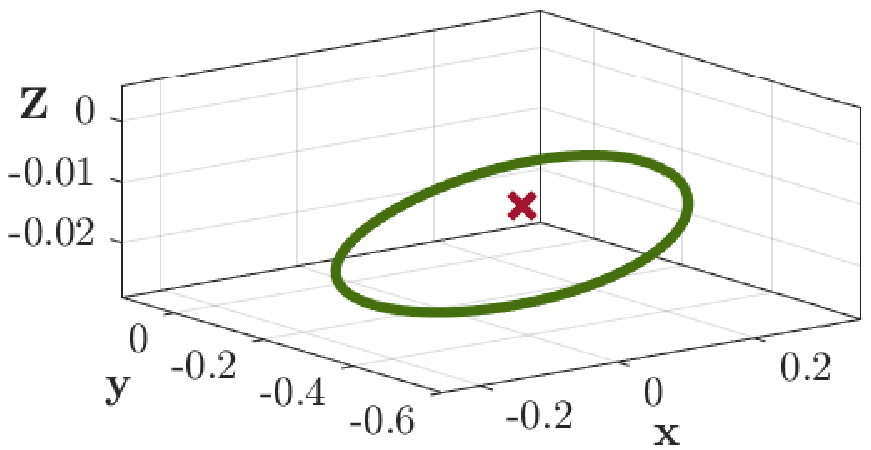}[1\baselineskip]{lc05}
    \subfigimg[width=0.49\columnwidth]{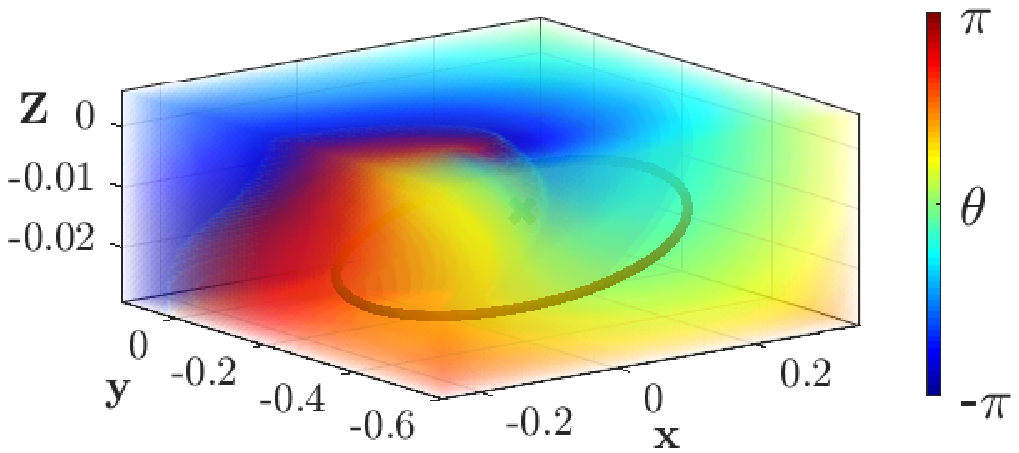}[1\baselineskip]{lc5}
    \subfigimg[width=0.48\columnwidth]{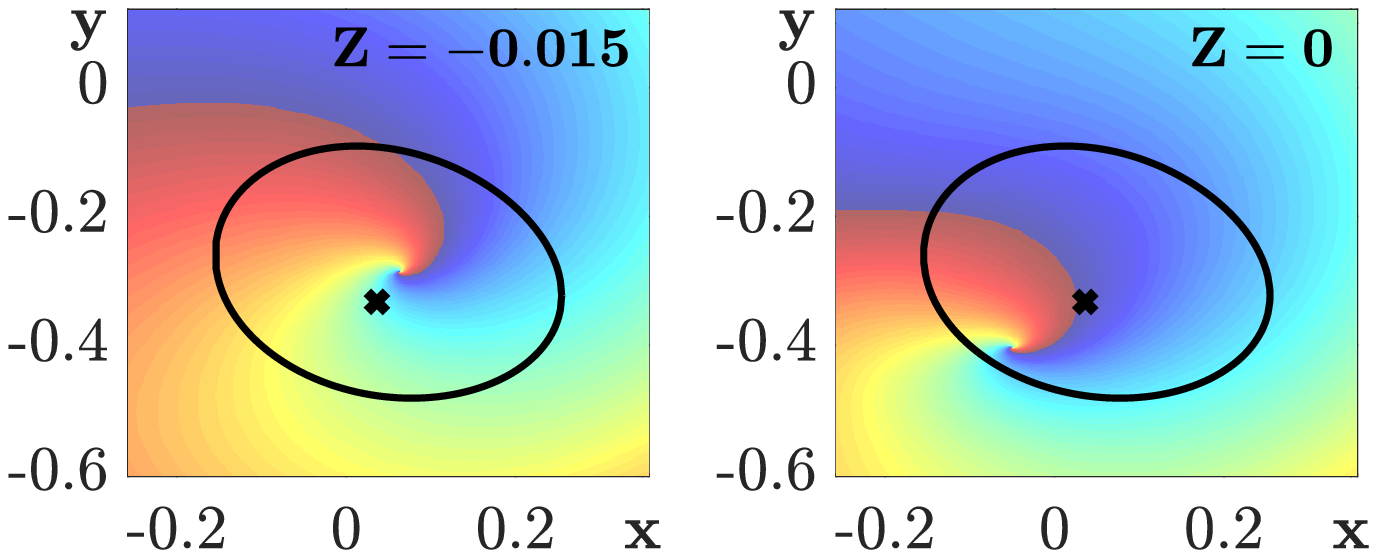}[1\baselineskip]{lc5_sections_z}
    \subfigimg[width=0.48\columnwidth]{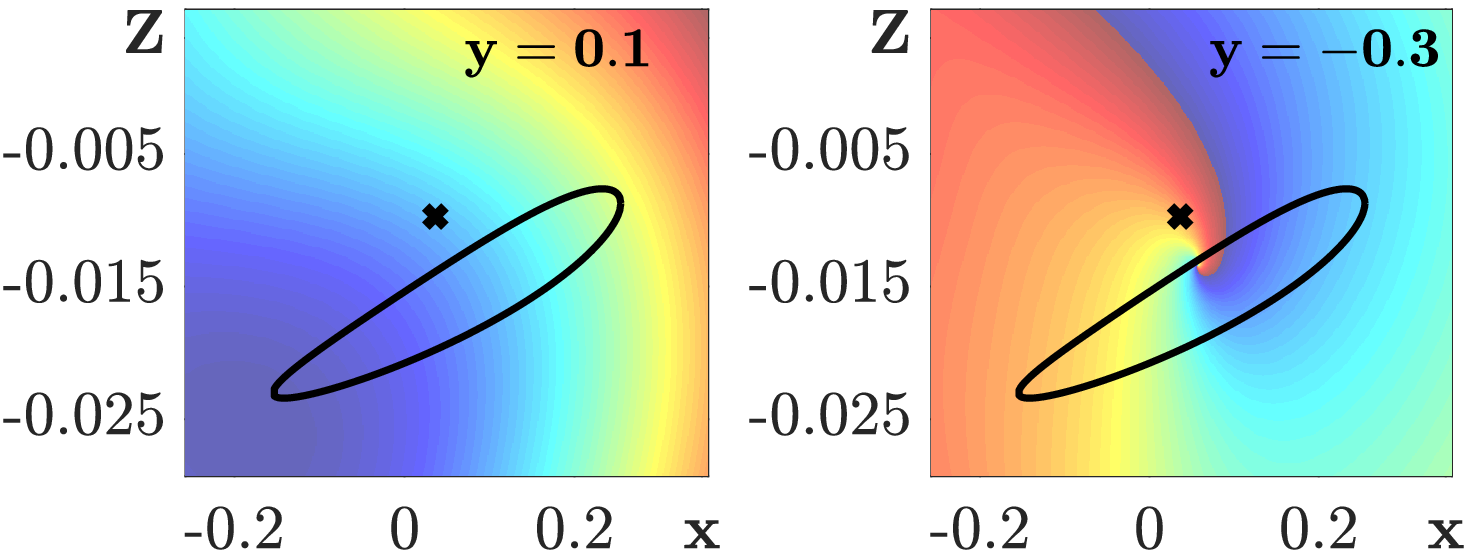}[1\baselineskip]{lc5_sections_y}
    \caption{(a) Phase space of the system for $\Omega=0.06$, $\eta=0.03$. x-points denote fixed points, green (thick) and red (light) curves denote stable and unstable limit cycles, respectively. (b) Isochron foliation of the 3D stable manifold of the LC. (c) Sections $\{Z=-0.015\}$, $\{Z=0\}$, $\{y=0.1\}$, $\{y=-0.3\}$ of (b). } 
    \label{5th}
\end{center}
\end{figure}

The structure of the isochrons in the phase space for each one of the aforementioned stable limit cycles are depicted in Figs. \ref{1st}-\ref{5th}, respectively. For the case shown in Fig. \ref{1st} $(\eta, \Omega)=(0.0005, 0.06)$ the isochron surfaces have a relatively simple form around the limit cycle and they are spiraling as approaching the saddle point; the phaseless set consists of the one-dimensional stable manifold of the saddle, close to which the isochrons accumulate. For the case depicted in Fig. \ref{2_nd} $(\eta, \Omega)=(0.0025, 0.06)$ the phaseless set includes, in addition to the one-dimensional stable manifold of the saddle, a coexisting unstable limit cycle, which significantly complicates the form of the isochrons. Further complications occur due to period-doubling, as shown in Fig. \ref{3rd} $(\eta, \Omega)=(0.01, 0.06)$. In this case, the phaseless set consists of the unstable limit cycle encircling one of the saddles, the two-dimensional stable manifold of this saddle, and the one-dimensional stable manifolds of the other two saddles. For values $(\eta, \Omega)=(0.02, 0.06)$ lying outside the region where period-doubling occurs, the phaseless set consists of the one-dimensional stable manifolds of the two saddles and the unstable node, and the form of the isochrons is relatively simplified, as shown in Fig. \ref{4th}. The form of the isochrons becomes again simple for $(\eta, \Omega)=(0.03, 0.06)$ for which the phaseless set consists solely of the one-dimensional stable manifold of the saddle, as shown in Fig. \ref{5th}.\\

[1] A. Dhooge, and W. Govaerts, Yu.A. Kuznetsov, H.G.E. Meijer, and B. Sautois, "New features of the software MatCont for bifurcation analysis of dynamical systems'', Math. Comput. Model Dyn. Syst. \textbf{14}, 147-175 (2008).

\newpage

\begin{figure}[h] 
\begin{center}
    \subfigimg[width=0.49\columnwidth]{lc02.eps}[1\baselineskip]{lc02}
    \subfigimg[width=0.49\columnwidth]{lc2.eps}[1\baselineskip]{lc2}
    \subfigimg[width=0.48\columnwidth]{lc2_sections_1x2_z}[1\baselineskip]{lc2_sections_z}
    \subfigimg[width=0.48\columnwidth]{lc2_sections_1x2_y}[1\baselineskip]{lc2_sections_y}
    \caption{(a) Phase space of the system for $\Omega=0.06$, $\eta=0.0025$. x-points denote fixed points, green (thick) and red (light) curves denote stable and unstable limit cycles, respectively. (b) Isochron foliation of the 3D stable manifold of the LC. (c) Sections $\{Z=-0.05\}$, $\{Z=0.05\}$, $\{y=0.1\}$, $\{y=-0.3\}$ of (b). } 
    \label{2_nd}
\end{center}
\end{figure}

\end{document}